\documentclass[prb,floatfix,superscriptaddress,nofootinbib,nolongbibliography]{revtex4-2}

\usepackage{subfigure}
\usepackage{xr-hyper}
\usepackage{hyperref}
\usepackage{svg}
\hypersetup{
     colorlinks   = true,
     citecolor    = red}
\usepackage{xr}
\usepackage{amsmath,amssymb}
\usepackage{bm}
\usepackage{gensymb,euscript}
\usepackage[normalem]{ulem}
\usepackage{euscript}
\usepackage{xspace}
\usepackage{xfrac}
\usepackage{enumerate}
\usepackage{braket}
\usepackage{float,fancyvrb}
\usepackage[T1]{fontenc}
\usepackage{lmodern}
\graphicspath{{./Figures/}}
\usepackage{wrapfig}
\usepackage{scrextend}
\usepackage{comment}
\usepackage{url}
\usepackage{amssymb}
\usepackage{xcolor} 
\usepackage{lineno} 
\usepackage{mathrsfs}
\usepackage[scr=zapfc, scrscaled=1.15]{mathalfa}
\usepackage{lineno}
\usepackage{subfloat}
\usepackage{float}
\usepackage{tabularx,booktabs}
\usepackage{ORCIDinREVTeX}

\newcommand{\uu}[1]{\ensuremath{\, \mathrm{#1}}} % units
\newcommand{\ja}[1]{\textcolor{blue}{#1}}
\usepackage[colorinlistoftodos]{todonotes}

\makeatletter
\newcommand*{\addFileDependency}[1]{% argument=file name and extension
  \typeout{(#1)}
  \@addtofilelist{#1}
  \IfFileExists{#1}{}{\typeout{No file #1.}}
}
\makeatother

\newcommand*{\myexternaldocument}[1]{%
    \externaldocument{#1}%
    \addFileDependency{#1.tex}%
    \addFileDependency{#1.aux}%
}
\myexternaldocument{supplement}
\begin{document}

%Here are some title suggestions
\title{Control of relaxation properties of a macroscopic nuclear spin ensemble}

\author{J\'anos \'Ad\'am}\orcid{0000-0003-0255-7606}
\affiliation{Department of Physics, Boston University, 590 Commonwealth Ave., Boston, MA 02215, USA}
\affiliation{Present address: 
IQM Quantum Computers, Espoo 02150, Finland}

\author{Andrew~J.~Winter}\orcid{0000-0002-6070-3723}
\affiliation{Department of Physics and Astronomy, Johns Hopkins University, Baltimore, Maryland 21218, USA}

\author{Deniz Aybas}
\orcid{0000-0002-0392-5979}
\affiliation{Department of Physics, Bilkent University, Ankara 06800, Turkey}

\author{Dmitry Budker} 
\orcid{0000-0002-7356-4814}
\affiliation{Johannes Gutenberg-Universit{\"a}t Mainz, 55122 Mainz, Germany}
\affiliation{Helmholtz Institute Mainz, 55099 Mainz, Germany}
\affiliation{GSI Helmholtzzentrum für Schwerionenforschung GmbH, 64291 Darmstadt, Germany}
\affiliation{Department of Physics, University of California, Berkeley, CA 94720-7300, United States of America}

\author{Derek F. Jackson Kimball} 
\orcid{0000-0003-2479-6034}
\affiliation{Department of Physics, California State University - East Bay, Hayward, California 94542-3084, USA}

\author{Arne Wickenbrock}
\affiliation{Johannes Gutenberg-Universit{\"a}t Mainz, 55122 Mainz, Germany}
\affiliation{Helmholtz Institute Mainz, 55099 Mainz, Germany}
\affiliation{GSI Helmholtzzentrum für Schwerionenforschung GmbH, 64291 Darmstadt, Germany}

\author{Alexander~O.~Sushkov} 
\orcid{0000-0001-8895-6338}
\affiliation{Department of Physics and Astronomy, Johns Hopkins University, Baltimore, Maryland 21218, USA}
\affiliation{Department of Physics, Boston University, 590 Commonwealth Ave., Boston, MA 02215, USA}

\date{\today}

\begin{abstract}
Macroscopic spin ensembles in solids are powerful platforms for quantum sensing and precision metrology. A key challenge is controlling the nuclear spin population relaxation time $T_1$, which can become prohibitively long at cryogenic temperatures due to phonon freeze-out. We demonstrate optical control of the $T_1$ relaxation time of the $^{207}$Pb nuclear spin ensemble in lead-containing ferroelectric crystals PbTiO$_3$ (PT) and (PbMg$_{1/3}$Nb$_{2/3}$O$_3$)$_{2/3}$-(PbTiO$_3$)$_{1/3}$ (PMN-PT). Using X-band electron paramagnetic resonance (EPR) spectroscopy at 10 K, we characterize light-induced paramagnetic centers created by 405 nm laser illumination. In PT, we observe paramagnetic Pb$^{3+}$ centers and their hyperfine interaction with nearby nuclear spins. 
In PMN-PT, we identify two populations: isotropic Pb$^{3+}$ centers and anisotropic Ti$^{3+}$ centers occupying $d$-orbitals, with spin number densities of $(2.5 \pm 1.0) \times 10^{17}$\,cm$^{-3}$ and $(4.1 \pm 1.7) \times 10^{17}$\,cm$^{-3}$, respectively. Power-dependent EPR measurements enable extraction of spin relaxation times. We investigate the ionization and recombination dynamics of these transient paramagnetic centers. Using saturation-recovery nuclear magnetic resonance, we demonstrate that laser illumination reduces the $^{207}$Pb nuclear $T_1$ by approximately a factor of two, from $(17 \pm 2)$\,s to $(7 \pm 1)$\,s at 4.6\,MHz, and from $(1550 \pm 40)$\,s to $(850 \pm 70)$\,s at 40\,MHz. We develop a model relating the nuclear relaxation rate to the density of photoinduced paramagnetic centers. This optical control of nuclear spin relaxation provides a pathway toward accelerated thermal polarization and dynamic nuclear polarization in solid-state NMR-based precision measurements, including searches for axion-like dark matter.
\end{abstract}
\maketitle

\hypersetup{linkcolor=blue}
%\linenumbers

\section{Introduction}

Spin ensembles are used and studied in many fields, such as medical imaging, materials science, quantum science, and fundamental physics~\cite{Degen2017,Safronova2018,Kuenstner2026}. Nuclear spins in solids are embedded in a host lattice, which acts as a thermal bath, with the corresponding relaxation time $T_1$. The dominant energy exchange mechanism between the spin ensemble and the lattice is usually phonon-induced relaxation~\cite{Abragam1961}. However, at low temperatures, the spin-lattice relaxation can become extremely slow, due to phonon freeze-out~\cite{Bloembergen1948}. Decoupling a quantum system from a thermal bath may seem advantageous since it permits coherent manipulation. However, it can be a bottleneck for efficient quantum state initialization~\cite{Bienfait2016}. For example, in nuclear magnetic resonance (NMR) spectroscopy and imaging, fast $T_1$ relaxation is often desirable to allow frequent experiment repetition and averaging~\cite{Ernst1990a,Levitt2008}. Controlling the $T_1$ relaxation is therefore an important tool when working with spin ensembles. Examples of approaches to such control include optical pumping, electrical initialization, and coupling to a resonant cavity~\cite{Bienfait2016,Albanese2020}. 

Our work focuses on controlling nuclear spin relaxation by manipulating a bath of paramagnetic impurities~\cite{Goldman1965}. This paramagnetic spin bath creates a fluctuating magnetic field that accelerates nuclear spin relaxation. Such control has been demonstrated in medicine and chemistry, where the goal is to enhance NMR spectroscopy signals of organic molecules using dynamic nuclear polarization (DNP)~\cite{VanKesteren1985,Diller2007,Dommaschk2015}. This approach is also used in magnetic resonance imaging, where contrast agents are widely employed~\cite{Michalak2011}.

A key motivation for the present work is the prospect that a nuclear spin ensemble in a non-centrosymmetric solid can be used as a quantum sensor to search for ultralight axion-like dark matter~\cite{Budker2014,Aybas2021a}. For this reason, we focus on studying Pb-containing ferroelectric solids lead titanate PbTiO$_3$ (PT) and the solid solution PMN-PT with chemical formula $(\text{PbMg}_{1/3}\text{Nb}_{2/3}\text{O}_3)_{2/3}$--$(\text{PbTiO}_3)_{1/3}$. The $^{207}$Pb nuclear spins in these materials experience an oscillating torque, induced by the defining QCD interaction of the axion dark matter field~\cite{Graham2013}. This torque can be quantified by the Rabi frequency $\Omega_a = g_d a_0 E^*/\hbar$, where $g_d$ is the coupling constant, $a_0$ is the amplitude of the axion-like dark matter field, $\hbar$ is the reduced Planck constant, and $E^*$ is the effective electric field, which is calculated to be $340$~kV/cm for $^{207}$Pb in PMN-PT~\cite{Aybas2021a}. If the Larmor frequency is tuned to be near the Compton frequency of the axion, this torque induces transverse magnetization of the spin ensemble with amplitude $M_a = u M_0 \Omega_a T_2$, where $M_0$ is the equilibrium spin magnetization, $T_2$ is the spin coherence time, and $u$ is a dimensionless spectral factor that takes into account the inhomogeneous broadening of the spin ensemble and the detuning between the axion-like particle Compton frequency and the spin Larmor frequency~\cite{Aybas2021a}.

The Cosmic Axion Spin Precession Experiment (CASPEr-electric, further referred to as CASPEr-e) searches for this axion dark matter-induced transverse magnetization. The first generation CASPEr-e search used the $^{207}$Pb nuclear spin ensemble in the ferroelectric PMN-PT. The search excluded axion-like dark matter with electric dipole moment (EDM) interaction strength $g_d$ greater than $9.5 \times 10^{-4}$~GeV$^{-2}$ in a 1~MHz band centered around 39.65~MHz, corresponding to the mass range 162 to 166~neV~\cite{Aybas2021a}.

The dark matter-induced magnetization $M_a$ is proportional to the equilibrium spin magnetization $M_0$, thus $M_0$ should be maximized in order to generate the largest possible dark-matter-induced signal. This can be accomplished with thermal pre-polarization in a higher magnetic field or DNP. However, DNP requires paramagnetic spins in the lattice and thermal pre-polarization requires a high magnetic field dwell time that exceeds the $T_1$ relaxation time, which can be prohibitively long. In the present work, we explore how laser illumination creates transient paramagnetic centers, which can be used to perform DNP hyperpolarization, or to suppress the $T_1$ relaxation time of the spin ensemble, enabling thermal pre-polarization. 

There are several previous studies of photo-induced paramagnetic centers in Pb-based relaxor ferroelectrics, using cryogenic electron paramagnetic resonance (EPR) spectroscopy. The materials that have been studied include lead titanate (PT: PbTiO$_3$), lead zirconium titanate (PZT: Pb(Zr$_x$Ti$_{1-x}$)O$_3$), and lead lanthanum-zirconium titanate (PLZT: Pb$_{1-y}$La$_y$Zr$_{1-x}$Ti$_x$O$_3$)~\cite{Laguta2000,Warren1996,Warren1993,Warren1992}. In all three of these materials, upon light illumination at cryogenic temperatures, a resonance appeared, with the $g$-factor $g = 1.995$ that was close to that of the free electron. It was hypothesized that upon illumination a small fraction of the Pb$^{2+}$ ions of the crystal further ionize to form Pb$^{3+}$ paramagnetic centers, and the unpaired electron spins give rise to the EPR spectral peak. This transition was modeled with the isotropic Zeeman Hamiltonian. In the case of PZT and PLZT another transition was detected around the $g$-factor of 1.934 which was assigned to Ti$^{3+}$ centers and modeled with an axial anisotropic Hamiltonian. 

In Sec.~\ref{sec:setup} we describe the experimental setup. In Sec.~\ref{sec:EPR} we present our experimental findings and model the observed EPR spectrum. In Sec.~\ref{sec:density} we use the parameters extracted from the spectrum to estimate the number of paramagnetic centers and determine their relaxation times. In Sec.~\ref{sec:dynamics} we investigate the temporal dynamics of the centers. In Sec.~\ref{sec:T1} we present our nuclear relaxation measurements with and without illumination and provide a model that explains how nuclear $^{207}$Pb relaxation time $T_1$ is affected by optical illumination of PMN-PT. In Sec.~\ref{sec:conclusion} we discuss the significance of our results in the context of solid state NMR-based precision measurements.

\section{Experimental setup} 
\label{sec:setup}

\begin{figure}[t!]
\includegraphics[width=0.6\linewidth]{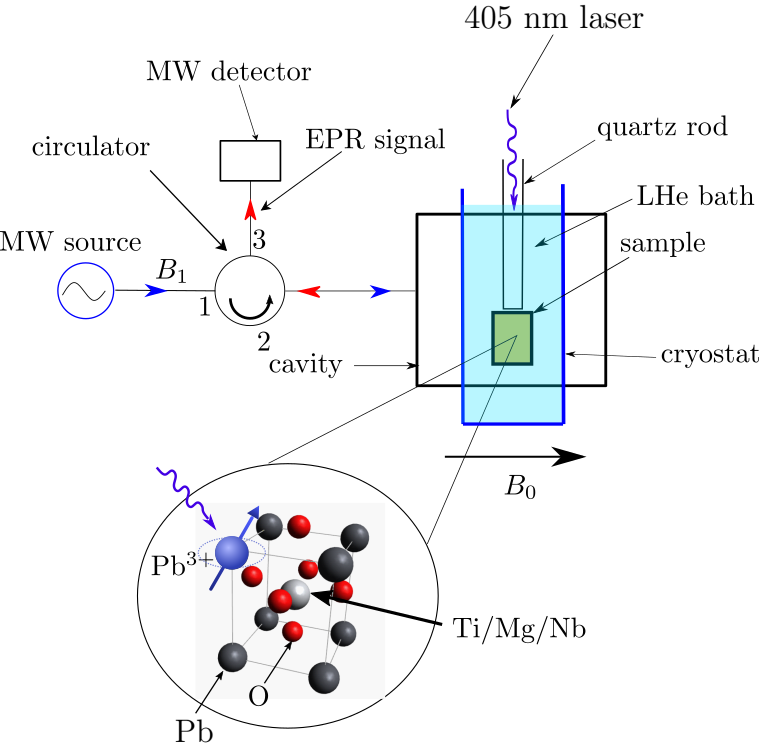}
\caption{Experimental setup schematic. The probe field (blue arrow) generated in the microwave source travels through a circulator and is coupled into a cavity. We place the sample (yellow rectangle) inside a temperature controlled cryostat. The sample is placed in the middle of the cavity (for PMN-PT we observed heavy losses therefore displaced the sample from the cavity center). A 405 nm laser light (purple wavy arrow) is coupled using a quartz rod. A bias magnetic field (\ensuremath{B_0}) is applied. After illumination paramagnetic centers are excited; this changes the impedance of the cavity which results in reflections (red arrow). The reflected signal travels through the circulator and gets absorbed in the microwave detector.\\
Inset: Schematic representation of PMN-PT, a cubic perovskite-like crystal. Pb atoms occupy the eight corner sites of the cube, oxygen atoms are located at the centers of each face, either Mg, Ti, or Nb resides at the body-centered position.\\
A wavy purple arrow indicates that the 405 nm laser acts on a corner Pb atom, inducing its conversion into a paramagnetic Pb center. Source of the lattice structure: Materials Project.}
\label{fig:setup}
\end{figure}

In our experiments we used the X-band Bruker Elexsys 9.4~GHz EPR spectrometer. The sample was placed in the Oxford Instruments liquid helium cryostat, with the base temperature of 10~K. The cryostat was positioned inside the EPR spectrometer cavity, Fig.~\ref{fig:setup}. Laser light at 405 nm was delivered to the cryostat via an optical fiber which was coupled to a quartz rod that guided the light to the sample.
The light power incident on the sample was measured to be $\approx2$~mW.

Our EPR spectroscopy measurements followed standard continuous-wave X-band procedures: for each measurement, the cavity was tuned with the sample in place to maximize absorbed microwave power and minimize reflected power at the detector, after which the static magnetic field was swept through resonance while a small modulation field was applied and the lock-in-detected first derivative of the absorption signal was recorded.
The presence of the sample changed the cavity resonance. Due to the high dielectric losses in PMN-PT it was difficult to tune the cavity unless the crystal was positioned away from the center of the cavity. Therefore, we placed it 15~mm away from the center (the length of the cavity is 40~mm). The PT sample was much less lossy, and therefore was placed in the middle of the cavity. After the cavity was tuned, the magnetic field sweep was initiated. An EPR transition at a certain magnetic field changes the cavity impedance and causes reflection of the incident microwave tone, which is detected by the microwave detector. The bias magnetic field was modulated with amplitude $\alpha = 2$~G and the detector signal modulation at the same frequency was recorded as the EPR signal.

\section{Analyzing EPR spectroscopy data for PT and PMN-PT samples}
\label{sec:EPR}

In order to establish our EPR spectroscopy protocol, our first experiments were performed with a PT crystal, with the goal of reproducing the results from Ref.~\cite{Warren1996}. In addition, quantitative results extracted from the PT crystal, especially the relaxation times, proved to be useful for estimating the PMN-PT relaxation times, as described below. 

\begin{figure}[h!]
  \centering
  \subfigure[]{
    \includegraphics[width=0.6\linewidth]{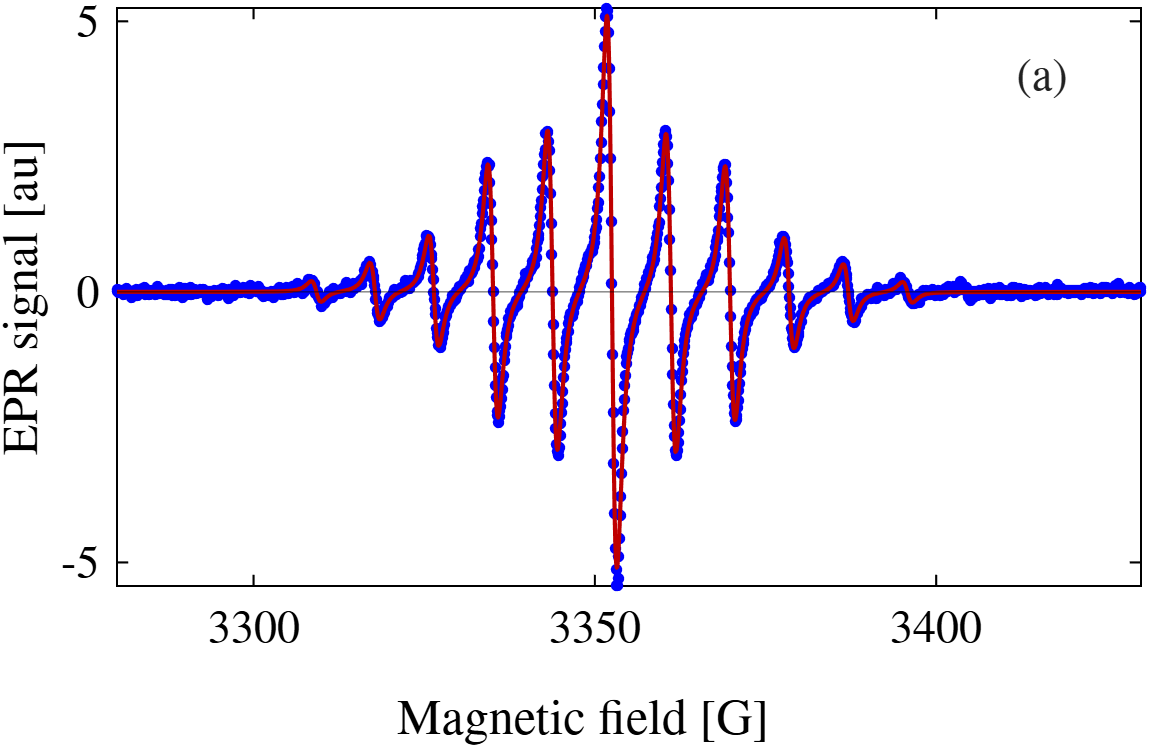}
    \label{fig:PTspec}
    }
  \subfigure[]{
  \includegraphics[width=0.6\linewidth]{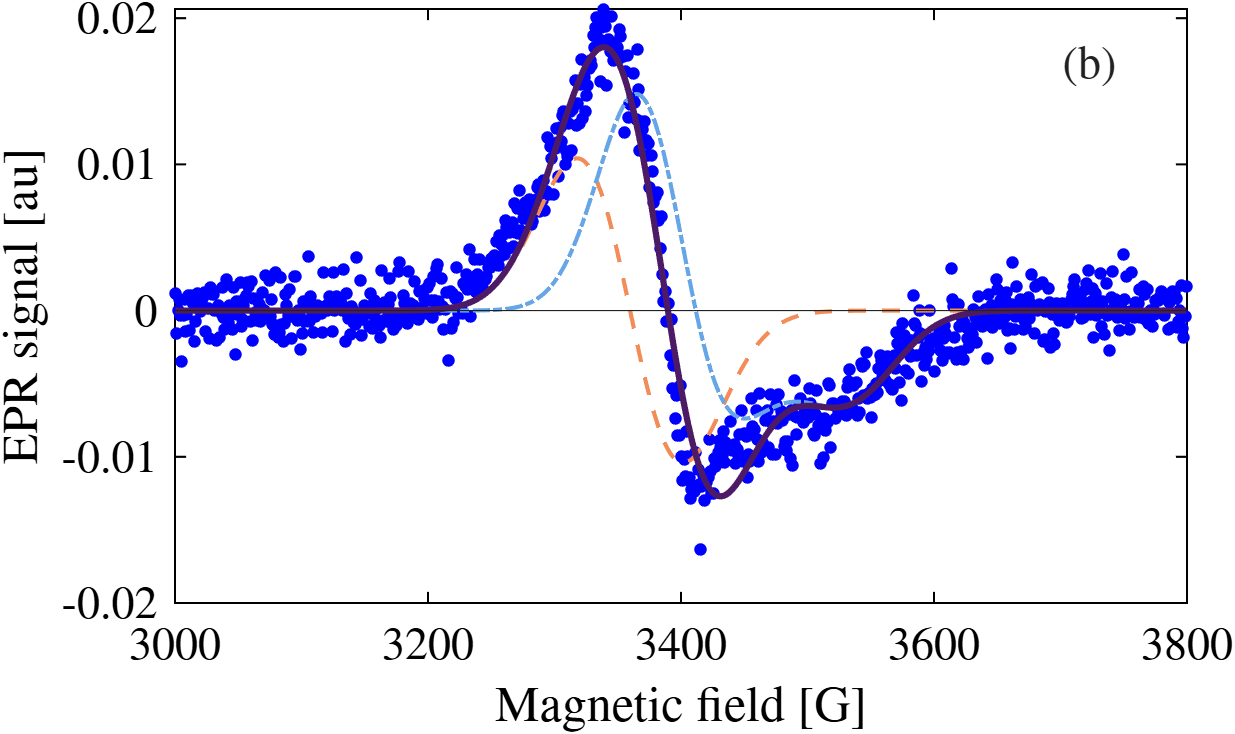}    
  \label{fig:PMNPTspec}    
    }
\caption{(a) The EPR spectrum of PT crystal at 10\,K and 0.2 mW (30\,dB attenuation) illuminated by 405 nm laser (blue dots). The signal belongs to the electron spins of $^{204}$Pb, $^{206}$Pb and $^{208}$Pb paramagnetic centers which have no nuclear spin. The central line is associated with the electron spin Zeeman interaction of the $I=0$ paramagnetic center, the satellite lines on each side arise from the super-hyperfine interactions of the electron spin with the nuclear spins on neighboring atomic lattice sites. We model the spectrum with the superposition of eleven Lorentzian lineshapes (red solid line). (b) the EPR spectrum of PMN-PT crystal at 10\,K and 1.3 mW (22\,dB attenuation) illuminated by 405 nm laser (blue dots) with background subtracted. We model the spectrum with a superposition (purple solid line) of signals from two spin populations: Pb$^{3+}$ (orange dashed line), Ti$^{3+}$ (light blue dashed line).
}
\label{fig:spec}
\end{figure}

The measured EPR spectrum of the PT crystal after laser illumination is shown in Fig.~\ref{fig:PTspec}. We observe the electron spins of the light-induced transient Pb$^{3+}$ paramagnetic centers~\cite{Warren1996}. This part of the spectrum corresponds to the centers located at the $^{204}$Pb, $^{206}$Pb and $^{208}$Pb (total abundance 78.9\%) isotopes that have no nuclear spin. The remaining stable isotope, $^{207}$Pb, has nuclear spin $I = 1/2$, and the hyperfine interaction displaces the corresponding EPR spectral features to much higher magnetic fields: 5520~G and 11220~G at 9.42~GHz EPR frequency~\cite{Warren1996}. 
In the present study we restrain our attention to the EPR spectra arising from the spinless Pb isotopes.
\renewcommand{\thesubfigure}{}

The central resonance line is associated with the electron spin Zeeman interaction of the $I = 0$ paramagnetic center. The satellite lines on each side of the central transition arise from superhyperfine interactions of the electron spin with the nuclear spins on neighboring atomic lattice sites~\cite{Warren1996}. We model the PT EPR spectrum with a superposition of eleven Lorentzian lineshapes: one at the center and five on each side. The central line is located at the magnetic field that corresponds to the g-factor $g = 2.001\pm0.005$, indicating isotropic magnetic Zeeman interaction corresponding to the paramagnetic center Pb$^{3+}$ occupying the $s$-orbital. 
All the Lorentzian components have the same linewidth (in magnetic field units): $\Delta_{\text{PT}} = (1.17 \pm 0.01)\,$G. Using this linewidth we can place a lower bound on the electron spin relaxation time: $\tau_2 > (\gamma \Delta_{\text{PT}})^{-1} = (48 \pm 0.5)\,$ns, where $\gamma$ is the free electron gyromagnetic ratio.
The uncertainty in the value of the $g$-factor is dominated by the systematic uncertainty of the value of the applied magnetic field in the EPR spectrometer. This magnetic field was separately calibrated using the EPR spectrum of the TEMPO free radical, whose central resonance is at $g = 2.006$~\cite{Gomez-Vidales2013,Cooper2010}.

Let us shift focus to the EPR spectrum of the PMN-PT crystal. In the absence of optical illumination, the spectrum consists of shallow broad lines (with feature widths much greater than $100\,$G) and linear drifts, we treat this as background and subtract it from the illuminated spectra in the following. Similar features appear with illumination but above cryogenic temperatures. The EPR spectrum of PMN-PT single crystal upon illumination after background subtraction is shown in Fig.~\ref{fig:PMNPTspec}. The spectrum shows a resonance line with asymmetric features. 
To model the shape of the observed PMN-PT EPR spectrum, we consider two distinct populations of paramagnetic centers~\cite{Warren1992,Warren1993,Warren1996,Laguta2000}. The first type of paramagnetic center is the isotropic Pb$^{3+}$, as observed in PT. The second type is the anisotropic Ti$^{3+}$ center. Such centers were not observed in PT, but previous studies of PZT and PLZT materials identified the existence of Ti$^{3+}$ centers. 

We model the magnetic Zeeman interaction of the light-induced paramagnetic centers with the Hamiltonian
\begin{equation}
    H = \mu_B \sum_{i = \{\text{Pb},\text{Ti}\}} g_i S_i^zB + \sum_{i = \{\text{Pb},\text{Ti}\}} \sum_j A_i I_j^z S_i^z,
    \label{eqn:Ham} 
\end{equation}
where the index $i$ indicates whether a variable belongs to the Pb or Ti population, $j$ runs over the nuclear spins, $\mu_B$ is the Bohr magneton, $B$ is the magnetic field, $S_i^z$ and $I_j^z$ are the projections of the electronic and nuclear spin operators along the direction of the magnetic field and $g_i$ is the $g$-factor and $A_i$ is the hyperfine coupling.

The first term describes the Zeeman interaction between the electronic spin and the external magnetic field. 
We use the magnetic field direction as the quantization axis. Based on our EPR results with PT sample, and Ref.~\cite{Warren1996}, the Pb$^{3+}$ centers have an isotropic g-factor, $g_{\text{Pb}}$.
However, the extra electron of the Ti$^{3+}$ paramagnetic center occupies a $d$-orbital, therefore its g-factor is anisotropic~\cite{Warren1992,Warren1993}.
Given the approximate uniaxial symmetry of the PMN-PT lattice, we model this as $g_{\text{Ti}}(\theta) = \sqrt{g_\parallel^2 \cos^2\theta + g_\perp^2 \sin^2\theta}$ where $\theta \in [0,\pi/2]$ is the angle between the principal crystal axis and the magnetic field and $g_\perp$, $g_{\parallel}$ are the transverse and longitudinal $g$-factors~\cite{Sands1955,Abragam1970}. Our PMN-PT sample is a single crystal, however there is intrinsic perovskite lattice disorder associated with the PMN-PT solid solution as well as with the ferroelectric domain structure. Therefore the direction of the principal axis is randomized throughout the macroscopic crystal. To obtain the EPR spectrum of the entire ensemble of paramagnetic centers, we use the powder-like orientational averaging approximation, integrating over the angle $\theta$, App.~\ref{app:B}. The result is the powder lineshape that describes the inhomogeneous distribution of spin resonance magnetic fields in an ensemble of randomly-oriented crystallites:
\begin{equation} 
    f_p(B) = \frac{1}{B^2}\frac{B_\parallel B_\bot ^2}{\sqrt{(B_\parallel^2-B_\bot^2)(B^2 - B_\bot^2)}},
    \label{eqn:2ddNdB}
\end{equation}
where $B_\bot = \hbar \omega'/\mu_B g_\bot$, $B_\parallel = \hbar \omega'/\mu_B g_\parallel$. We follow the standard practice of EPR spectroscopy to use the magnetic field $B$ as the independent variable, given a fixed microwave resonator angular frequency $\omega'$.

The second term in the Hamiltonian describes the hyperfine interaction between the paramagnetic impurities and nearby nuclear spins. We assume isotropic hyperfine interaction because the anisotropic components average out in the randomized powder sample. 
The PT spectra showed resolved superhyperfine satellites. In contrast, in PMN-PT the corresponding hyperfine/superhyperfine structure is unresolved because each center is coupled to a dense bath of $^{93}$Nb nuclei ($I=9/2$, 100\% abundance), producing many overlapping transitions that appear experimentally as inhomogeneous broadening.
Therefore our spectral lineshape model for PMN-PT does not explicitly include the hyperfine interaction parameters, but treats it as inhomogeneous broadening, with the Gaussian lineshape~\cite{Bales1998}.

The spectral lines assigned to the Pb$^{3+}$ paramagnetic centers are modeled with the Gaussian lineshape
\begin{equation}
F_{P}(B) =\frac{1}{\sqrt{2\pi} \Delta_P} \exp\Bigg [-\frac{(B-B_P)^2}{2 \Delta_P ^ 2 } \Bigg ],
\label{eq:PbGB}
\end{equation}
where $B$ is the scanned static magnetic field, $B_P = \hbar \omega/\left(\mu_B g_{\text{Pb}}\right)$ is the line center, and $\Delta_{\text{P}}$ is the inhomogeneous linewidth, App.~\ref{app:B}. 

The spectral lines assigned to the Ti$^{3+}$ paramagnetic centers are also affected by the inhomogeneous broadening, due to the unresolved hyperfine structure, but the anisotropic $g$-factor in the first term of Eq.~\eqref{eqn:Ham} is an equally important broadening mechanism. We model the corresponding lineshape as a convolution of the Gaussian line and the powder lineshape $f_p$:
\begin{align}
F_{T}(B) = \int \frac{f_p(B')}{\sqrt{2\pi} \Delta_T} \exp\Bigg [-\frac{(B-B')^2}{2 \Delta_T ^ 2 } \Bigg ] dB',
    \label{eq:TiGB}
\end{align}
where $\Delta_{\text{T}}$ is the inhomogeneous linewidth, App.~\ref{app:B}. 

The EPR spectrometer introduces a modulation on top of the applied magnetic field, in order to reduce the deleterious effects of low-frequency noise. Therefore, the detected voltage is proportional to the derivative of the absorption lineshape with respect to magnetic field. With this in mind, our model for the EPR spectrum of Pb$^{3+}$ and Ti$^{3+}$ light-induced paramagnetic centers in PMN-PT is given by:
\begin{align}
    V=V_{Pb} + V_{Ti} &= A_P\frac{d F_{P}(B)}{dB} + A_T\frac{dF_{T}(B)}{dB}.
     \label{eq:39}
\end{align}
We find that this model is consistent with the main features of our experimental spectra, Fig.~\ref{fig:spec}, allowing us to extract best-fit values of the g-factors and inhomogeneous linewidths.

The best-fit value of the $g$-factor of Pb$^{3+}$ light-induced paramagnetic centers is $g_{\text{Pb}} = 2.001 \pm 0.005$. This is in agreement with previous studies on PT, PZT and PLZT. The best-fit anisotropic $g$-factor parameters of the Ti$^{3+}$ light-induced paramagnetic centers are: $g_\perp = 1.902 \pm 0.005$ and $g_\parallel = 1.987 \pm 0.005$. The best-fit inhomogeneous linewidths are $\Delta_{\text{Pb}} = 42 \pm 8$~G and $\Delta_{\text{Ti}} = 36 \pm 8$~G. As for the PT measurements, the $g$-factor uncertainties are dominated by the systematic uncertainty of the value of the applied magnetic field in the EPR spectrometer.

\section{Spin density and spin relaxation times} 
\label{sec:density}

\begin{figure}[h!]
  \centering
  \subfigure[]{
  \includegraphics[width=0.5\linewidth]{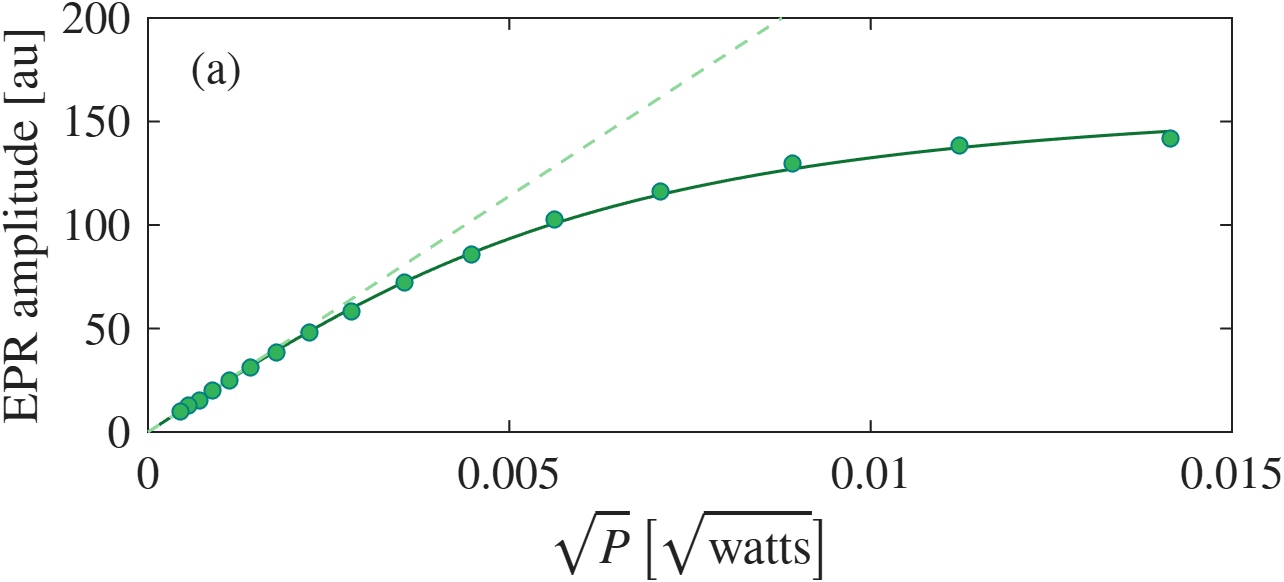}
  \label{fig:PTsat}
}
\subfigure[]{

  \includegraphics[width=0.5\linewidth]{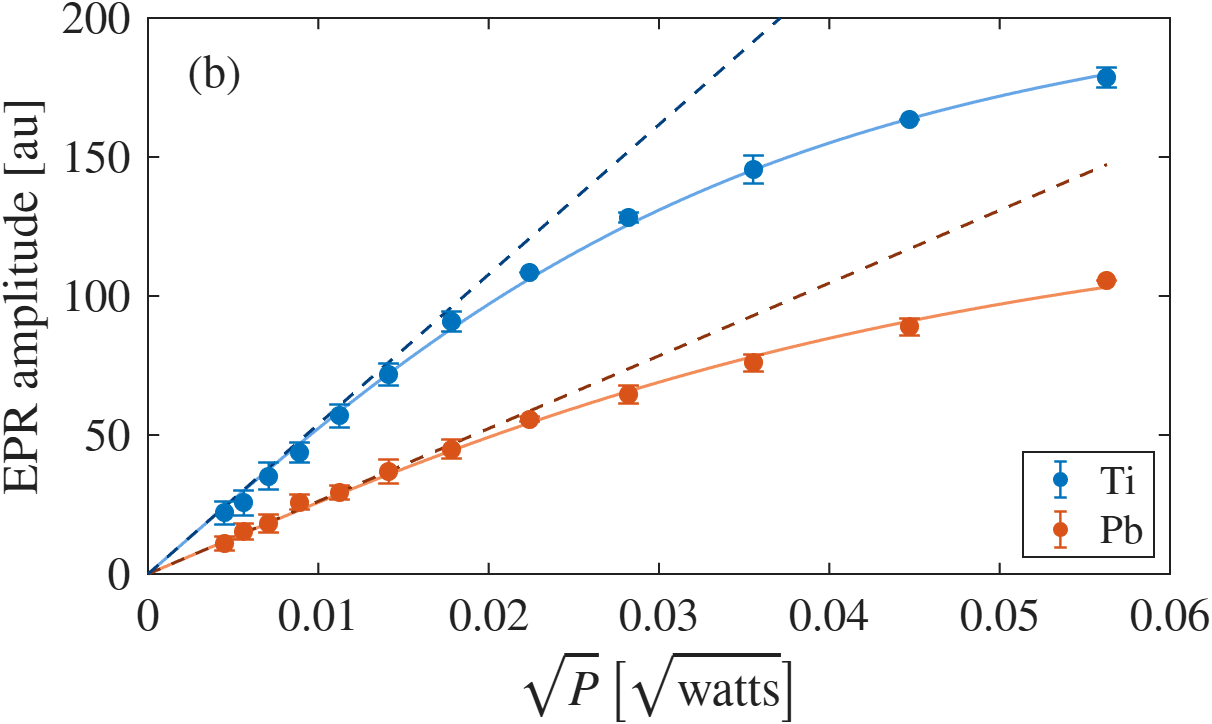}
  \label{fig:PMNPT_Pbsat}
}
\caption{
(a) Saturation curve of PT. The spectrum of PT has been measured at different microwave amplitudes and the area under the data has been integrated (green dots). The data has been fitted (dark green line) with Eq.~\eqref{eq:ampeq} where $a = 2.3 \times 10^4$, $b = 2 \times 10^4$.  The signal saturates at high amplitudes due to power broadening. The linear equation, without the power broadening denominator term is shown with light green dashed line.  
(b) Saturation curve of PMNPT. The spectrum of PMN-PT has been measured at different microwave powers, similar to (a). However, since the spectrum is the superposition of two different populations, direct integration is not possible. Therefore the fitted curves has been integrated and shown here (blue for Ti$^{3+}$ and orange for Pb$^{3+}$). 
The data has been fitted (blue and orange solid lines accordingly) and the purely linear part is also shown (blue and orange dashed lines). 
The coefficients for Ti$^{3+}$ are $a = 5.4 \times 10^3$, $b = 5.8 \times 10^2$ and for Pb$^{3+}$ are $a = 2.6 \times 10^3$, $b = 3.3 \times 10^2$.
}
\label{fig:sat}
\end{figure}
The amplitudes $A_P$, $A_T$ of the EPR spectral lines in Eq.~\eqref{eq:39} are proportional to the number of paramagnetic centers in the sample. In order to infer these numbers, we varied the microwave power that was used for EPR measurements and observed the saturation behavior of the spectra over a range of drive powers, Fig.~\ref{fig:sat}. We model the dependence of the amplitude on microwave power $P$ as follows:
\begin{align}
A(P) = \frac{a \sqrt{P}}{\sqrt{1 + b P}},
\label{eq:ampeq}
\end{align}
where $a$ is a parameter proportional to the paramagnetic center number density $n$, and $b$ is a parameter proportional to the product of the population and coherence spin relaxation times $\tau_1\tau_2$, App.~\ref{app:B}. The voltage detected by an EPR spectrometer is $\propto\sqrt{P}$, which appears in the numerator~\cite{Eaton2010}. The denominator describes the standard power broadening and saturation behavior of a two-level system driven by a near-resonant field $B_1$, giving the saturation factor $1/\sqrt{1 + \gamma^2 B_1 ^ 2 \tau_{1} \tau_{2}}$, where $B_1^2\propto P$ and $\gamma$ is the gyromagnetic ratio. The derivation of Eq.~\eqref{eq:ampeq} and the expressions for $a$, $b$ in terms of experimental parameters are detailed in App.~\ref{app:B}.
These expressions contain a number of parameters that are difficult to measure accurately. Therefore we calibrated the spectrometer using the TEMPO radical calibration sample containing $N_0 = 4\times10^{16}$ electron spins. We inferred spin number densities of the Pb$^{3+}$ and the Ti$^{3+}$ light-induced paramagnetic centers by rescaling the best-fit values of the corresponding $a$ parameters by the value of $a$ for the TEMPO calibration sample, with a correction for the value of the EPR cavity quality factor measured in both cases, App.~\ref{app:C}. We also used the best-fit values of the $b\propto\tau_1\tau_2$ parameter to estimate the spin relaxation timescales $\tau_1$, $\tau_2$. 

The measured EPR spectra of the PT sample under light illumination at $10\,$K correspond to the Pb$^{3+}$ spin number density $(1.33 \pm 0.25) \times 10^{17}$~cm$^{-3}$. Using the value $\tau_{2,P} \approx 48\,$ns, quoted in Sec.~\ref{sec:EPR}, we obtain the estimate $\tau_{1,P} = (160 \pm 25)~\mu$s.

The measured EPR spectra of the PMN-PT sample under light illumination at $10\,$K correspond to the Pb$^{3+}$ spin number density $(2.5 \pm 1.0) \times 10^{17}$~cm$^{-3}$ and the Ti$^{3+}$ spin number density $(4.1 \pm 1.7) \times 10^{17}$~cm$^{-3}$. Estimates of spin relaxation timescales in PMN-PT are complicated by the inhomogeneously-broadened EPR linewdiths. For the purposes of a rough estimate, we make the assumption that $\tau_{2,P}$ in PMN-PT is the same as in PT, which implies that in PMN-PT $\tau_{1,P} \approx (31 \pm 12)~\mu$s. Since we have no independent measurements of Ti$^{3+}$ spin relaxation times, we make no estimates for this spin ensemble.

\section{The ionization and recombination dynamics of the light-induced paramagnetic centers}
\label{sec:dynamics}

\begin{figure}[h!]
  \centering
  \subfigure[]{
    \includegraphics[width=0.6\columnwidth]{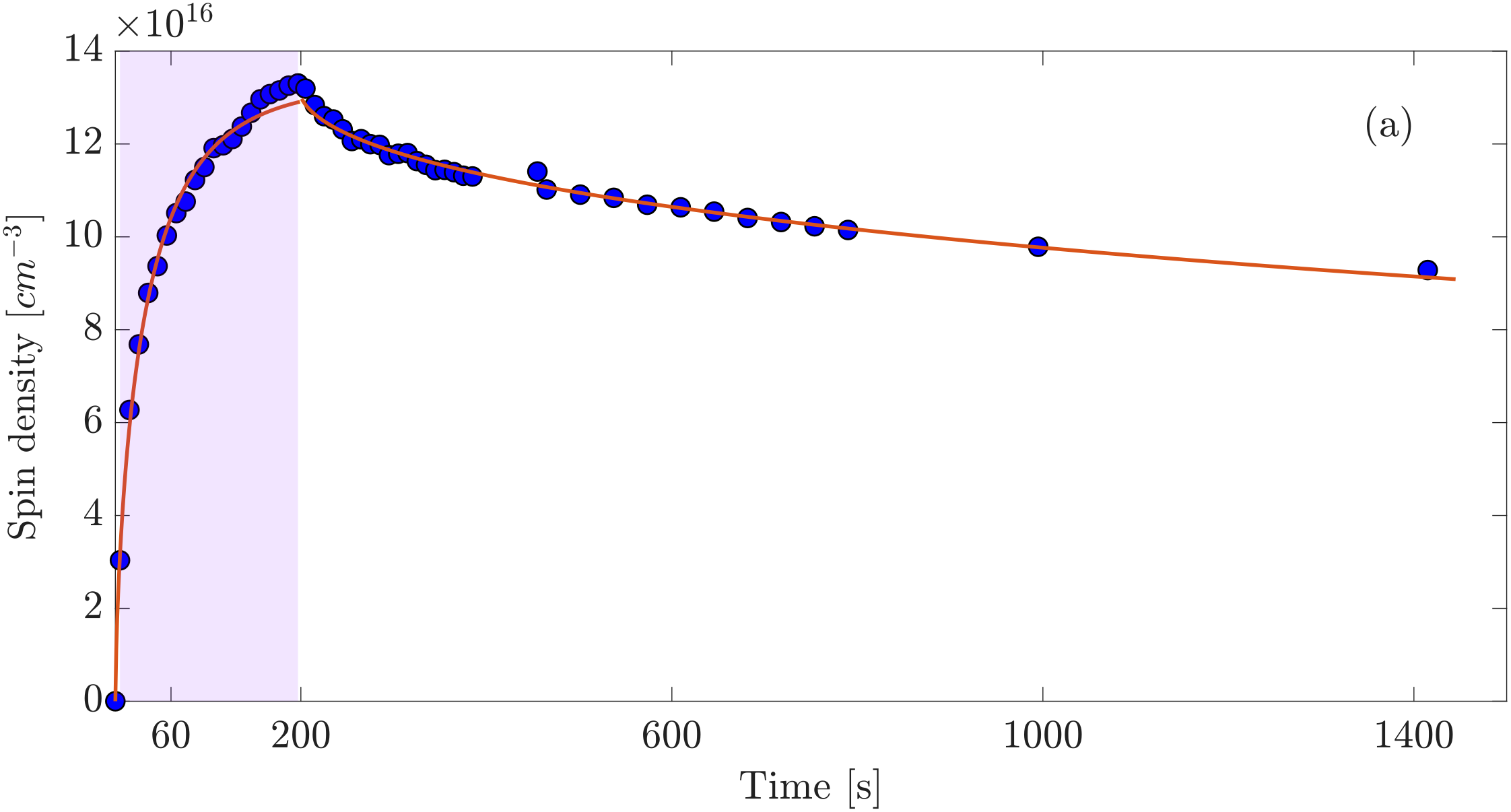}
    \label{fig:PTdecay}
    }

  \subfigure[]{
\includegraphics[width=0.6\columnwidth]{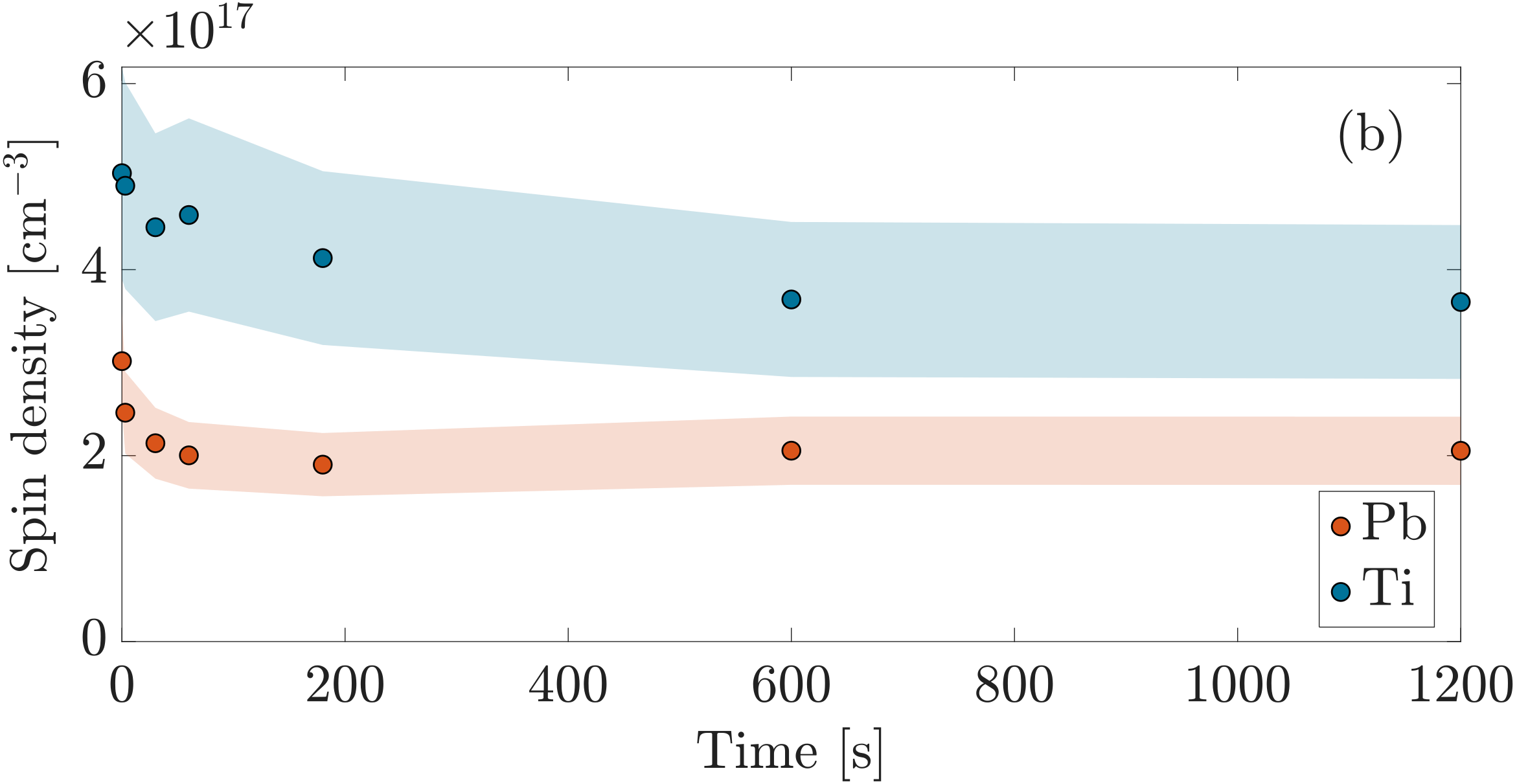}
    \label{fig:PMNPT_spec_decay}
    }

  \subfigure[]{
\includegraphics[width=0.6\columnwidth]{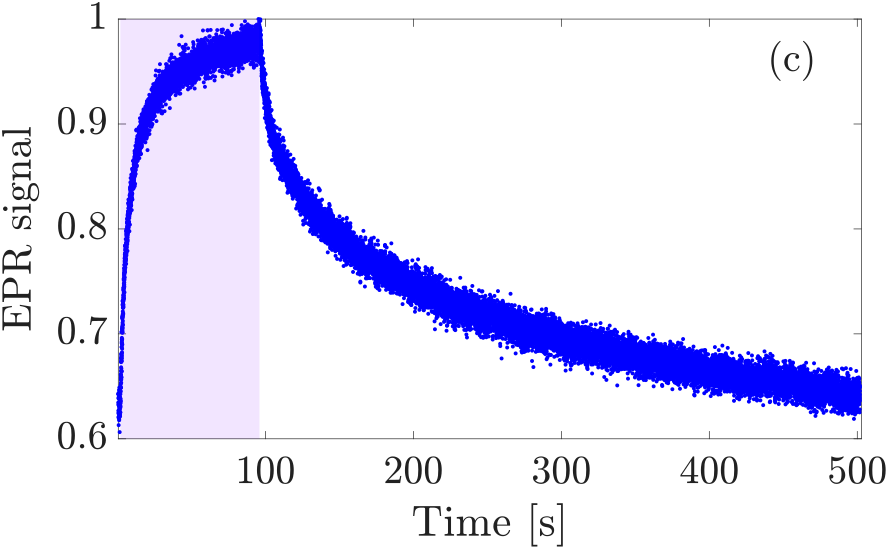}
    \label{fig:PMNPT_fixed_decay}
    }
\caption{The photo-ionization and recombination dynamics of light-induced paramagnetic centers.
	(a) Build-up and decay of paramagnetic centers in PT at 10\,K. The purple shaded area shows when the laser was on. 
	(b) The decay of densities of Pb$^{3+}$ and Ti$^{3+}$ centers in PMN-PT: the laser light was turned off at $t=0$. The spin densities were calculated based on the fits used in Eq.~\eqref{eq:39}. The shaded bands indicate the systematic uncertainty on the spin density, which is dominated by the uncertainty in the sample position within the cavity, App.~\ref{app:C}.
	(c) Time dependence of normalized PMN-PT signal at a fixed magnetic field. The magnetic field is chosen so that it corresponds to the maximum of the spectrum at 10\,K and 2\,mW illumination. The purple shaded area shows when the laser was switched on.} 
\label{fig:time}
\end{figure}

In the previous sections we presented our measurements of EPR spectra of PT and PMN-PT under steady-state optical illumination at $10\,$K. In the present section we investigate the time dependence of the optically excited centers. The dynamics are qualitatively similar for PT and PMN-PT: upon illumination there is a buildup of paramagnetic center density, and when the illumination is turned off the density decays due to recombination, Fig.~\ref{fig:PTdecay}. It is clear that the ionization and recombination dynamical timescales are on the order of tens to hundreds of seconds and the dynamics are not exponential~\cite{Bairavarasu2006}. Both the illumination time and intensity can be used to control the density of paramagnetic centers, which can persist in the samples for time scales much longer than 1000\,s.

For PT we recorded the spectra and integrated them at different times. The integral of the spectra is proportional to the spin density. Figure~\ref{fig:PTdecay} shows the time dependence of the spin density for a PT sample. The data show that the recombination is a slower process compared to the photo-excitation. 
We model the spin density dynamics with the stretched exponential time dependence $\Delta n(t)\propto \exp{(-(\beta t)^\kappa)}$. The best-fit values of the parameters $\beta,\,\kappa$ are listed in Tab.~\ref{tab:n}.
%------------------------------------------
\begin{table}[!hb]
    \centering
    \renewcommand{\arraystretch}{2.0}
    \setlength{\tabcolsep}{8pt}
    \fontsize{11pt}{11pt}\selectfont
\begin{tabular}{|c|c|c|}
\hline
 & $\kappa$ & $\beta$ (s$^{-1}$) \\
\midrule
laser on & $0.70 \pm 0.02$ & $(31 \pm 1 )\times 10^{-3}$ \\
\hline
laser off & $0.48 \pm 0.01$ & $(0.11 \pm 0.01) \times 10^{-3}$ \\
\bottomrule
\end{tabular}
\caption{\fontsize{10}{11}\selectfont
	The density dynamics of light-induced paramagnetic centers in a PT crystal: photo-ionization and recombination.}
\label{tab:n}
\end{table}
%------------------------------------------

Similarly to PT, we recorded PMN-PT spectra at a few data points. We fitted the spectra as in Sec.~\ref{sec:EPR} and extracted the spin densities. The decay of these spin densities for the two populations in PMN-PT is shown in Fig.~\ref{fig:PMNPT_spec_decay}.
For the PMN-PT sample, we also recorded the time dependence of the signal at a fixed static magnetic field (Fig.~\ref{fig:PMNPT_fixed_decay}).\footnote{We picked the magnetic field corresponding to the maximum of the spectrum.} This allows us to record the decay at more data points, but without access to the full spectrum we cannot distinguish between the decay rates of the two populations. We illuminated the sample for 100~seconds and recorded data for 400 more seconds after the illumination has been turned off. 
Both PMN-PT plots indicate that there are two processes governing the decay: a quick drop, followed by a slow decay---similarly to the phenomenon observed in Ref.~\cite{Bairavarasu2006}.

\section{Controlling nuclear spin relaxation with light-induced paramagnetic centers}
\label{sec:T1}

Nuclear spin population relaxation time $T_1$ can be extremely long in insulating solids at cryogenic temperatures~\cite{Waugh1988}. This can be deleterious for nuclear magnetic resonance (NMR) measurements, due to long polarization and averaging times. Due to slow nuclear spin-lattice relaxation at low temperature, nuclear $T_1$ is often limited by the fluctuating magnetic field produced by paramagnetic impurities in the lattice~\cite{Khutsishvili1969,Henrichs1984}. Our EPR measurements show that we can use optical illumination to control the density $n_s$ of light-induced paramagnetic centers. Therefore, we should be able to control the nuclear spin $T_1$. 

\begin{figure}[h]
  \centering
   \subfigure[]{
 \includegraphics[width=0.5\linewidth]{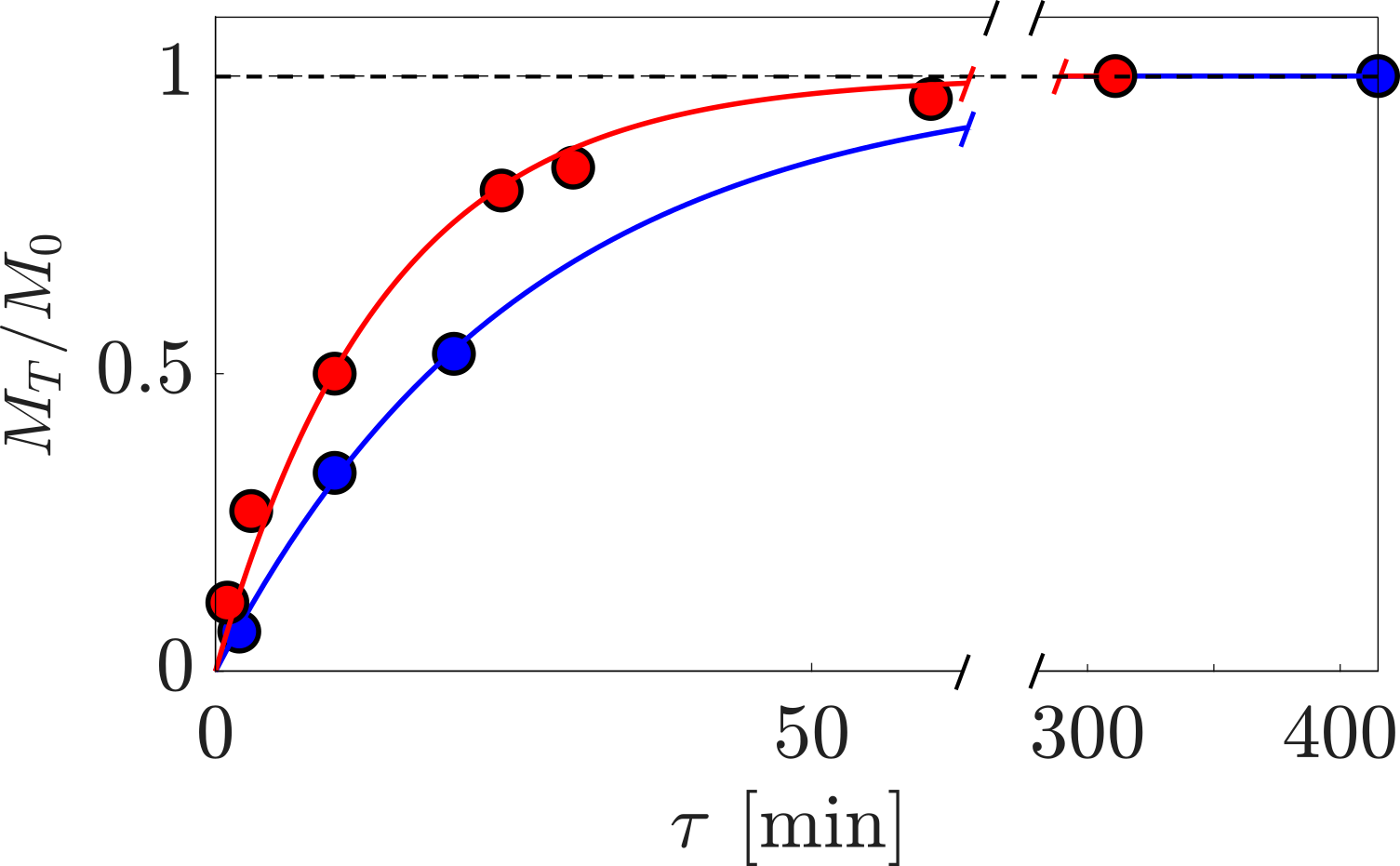}  
    \label{fig:PbSatRec}
    }
\subfigure[]{
 \includegraphics[width=0.5\linewidth]{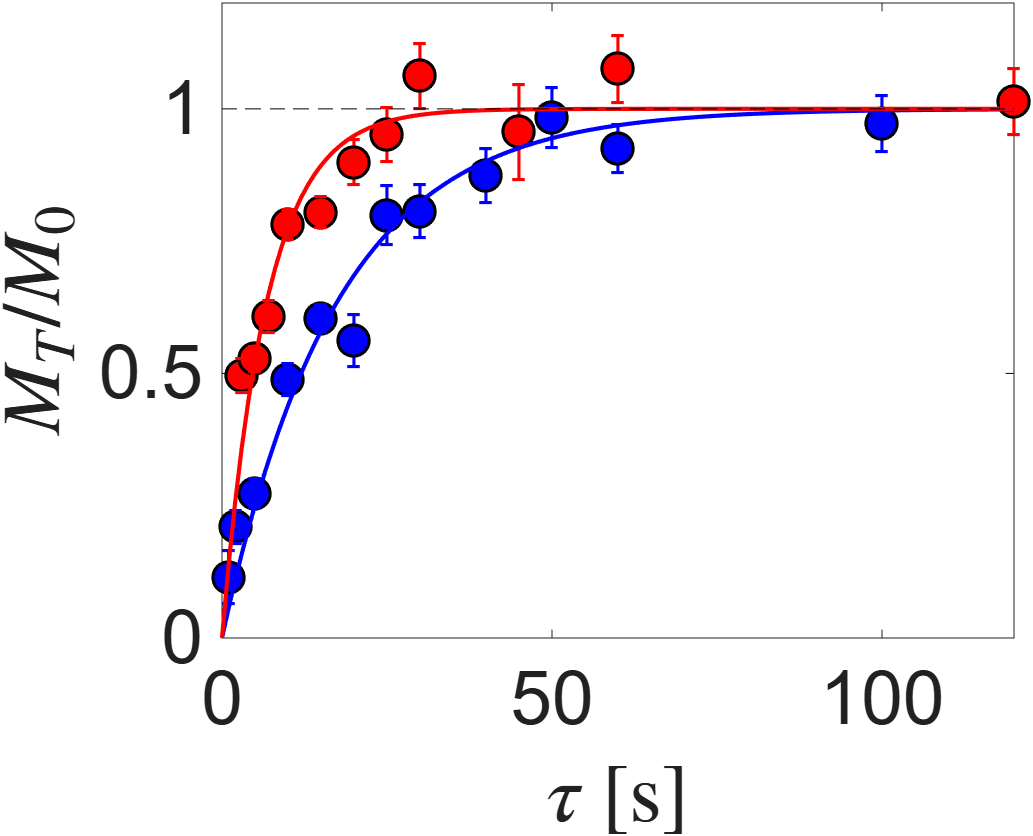}  
    \label{fig:PbSatRec}
    }
  \caption{Pb NMR saturation-recovery measurement of PMN-PT with light off (blue) and light on (red). The Y-axis shows transverse magnetization, normalized in units of maximal longitudinal magnetization. The X-axis shows the saturation recovery delay time $\tau$. (a) The measurements at 40\,MHz NMR frequency. (b) The measurements at 4.6\,MHz NMR frequency. 
}
\end{figure}

The details of our NMR experiments are described in Refs.~\cite{Aybas2021a,Adam2023}. The apparatus was different than that for the EPR measurements: the PMN-PT sample was immersed in a liquid helium bath inside the bore of a superconducting solenoid magnet, and a 405$\,$nm laser was coupled to the sample with an optical fiber. 
The nuclear spin population relaxation was measured using the saturation recovery experiment, and the data were modeled with exponential time dependence, Fig.~\ref{fig:PbSatRec}. Before illuminating the sample with the laser, we performed baseline relaxation measurements ``in the dark,'' extracting $T_1^{(d)}$. We then illuminated the sample with the laser, and repeated the saturation recovery experiment, extracting $T_1^{(\ell)}$. Our measurements demonstrate that laser illumination can control the value of the $^{207}$Pb nuclear spin relaxation time, Tab.~\ref{tab:T1}. 
%------------------------------------------
\begin{table}[!hb]
    \centering
    \renewcommand{\arraystretch}{2.0}
    \setlength{\tabcolsep}{8pt}
    \fontsize{11pt}{11pt}\selectfont
\begin{tabular}{|c|c|c|}
\hline
$^{207}$Pb NMR frequency (MHz) & $T_1^{(d)}$ (s) & $T_1^{(\ell)}$ (s) \\
\midrule
4.6 & $17 \pm 2$ & $7 \pm 1$ \\
\hline
40 & $1550 \pm 40$ & $850 \pm 70$ \\
\bottomrule
\end{tabular}
\caption{\fontsize{10}{11}\selectfont
	Measured values of $^{207}$Pb nuclear spin relaxation times in a PMN-PT crystal.}
\label{tab:T1}
\end{table}
%------------------------------------------

Let us consider a model of the physical mechanism for how the nuclear relaxation time is affected by paramagnetic centers. We consider the ensemble of $^{207}$Pb nuclear spins in our PMN-PT crystal. The frequency scale that corresponds to magnetic dipolar interactions between nearest-neighbor $^{207}$Pb spins is $J_d\approx 200\,$Hz~\cite{Mukhamedjanov2005a}. An important NMR line broadening mechanism is chemical shift anisotropy (CSA), which is $\approx 1000\,$ppm in our samples~\cite{Aybas2021a}.
We performed NMR measurements at two different magnetic fields, corresponding to $^{207}$Pb Larmor frequencies of $4.6\,$MHz and $40\,$MHz. At both frequencies the CSA dominates the dipolar linewidth and nuclear spin diffusion is suppressed. Therefore we consider direct relaxation of nuclear spins by a fluctuating electron spin of a paramagnetic center. If a nuclear spin is a distance $r$ away from an electron spin, the rate of nuclear spin flips is given by
\begin{align}
    w = \frac{C}{r^6} = \frac{2}{5} \frac{\hbar^2 \gamma_s^2 \gamma_n^2 S (S+1)}{r^6} \frac{\tau_{c}}{1+ \omega_n^2 \tau_{c}^2},
\end{align}
where $\gamma_s$ is the electron spin gyromagnetic ratio, $\gamma_n$ is the nuclear gyromagnetic ratio (in this case $^{207}$Pb), $S$ is the electron spin magnitude, $\tau_{c}$ is the electron spin correlation time, and $\omega_n$ is the nuclear Larmor angular frequency~\cite{Goldman1965}. For a typical nuclear spin, the distance $r_0$ to the closest electron spin can be estimated by half of the mean separation between electron spins: $r_0 = 1/(2 n_{s}^{1/3})$, where $n_s$ is the electron spin density. Therefore the total nuclear spin relaxation rate can be estimated as the integral over the lattice:
\begin{align}
1/T_1 \approx \int w n_{s} d^3\mathbf{r} = 4 \pi C n_{s} \int_{r_0}^\infty r^{-4} dr = \frac{4 \pi}{3} \frac{n_{s}C}{r_0^3} = \frac{64 \pi}{15} n_{s}^2 \hbar^2 \gamma_s^2 \gamma_n^2 S (S+1)\frac{\tau_{c}}{1+ \omega_n^2 \tau_{c}^2}.
\label{eq:T1}
\end{align}
If we use $n_s\approx 2\times10^{17}\,$cm$^{-3}$ (obtained from our EPR measurements), then our nuclear $T_1$ relaxation measurements are consistent with this model if we assume that the electron spin correlation timescale of $\tau_c\approx 10\,$ms. This appears to be much longer than the estimates of the electron spin population relaxation time $\tau_1\approx100\,\mu$s, based on our EPR measurements. However, the EPR measurements were performed at the temperature of $\approx10\,$K, whereas the NMR measurements were performed at the temperature of $4\,$K, and the spin population relaxation rates are known to depend strongly on temperature.

\section{Conclusions and Outlook} 
\label{sec:conclusion}

The ability to generate a controlled density of photoinduced paramagnetic centers in Pb-containing ferroelectric crystals provides a powerful route toward precision magnetic-resonance experiments~\cite{Bienfait2016}.  In the present work we studied the nature and dynamics of light-induced paramagnetic centers in PbTiO$_3$ and PMN-PT. We demonstrated control over $^{207}$Pb nuclear spin $T_1$ relaxation time in PMN-PT. This capability can be used to accelerate the buildup of thermal polarization at cryogenic temperatures, where $T_1$ can otherwise become prohibitively long. More broadly, paramagnetic centers offer a path to enhancing nuclear-spin polarization via dynamic nuclear polarization (DNP). Such nuclear hyperpolarization would increase the sensitivity of searches for new fundamental physics, including the CASPEr search for the QCD axion~\cite{Budker2006,Budker2014,DeMille2017,Wu2019a,Garcon2019,Aybas2021a,Aybas2021b,Sushkov2023}. 

\section*{Acknowledgements}
This work was supported by the U.S. National Science Foundation CAREER grant PHY-2145162, the U.S. Department of Energy grant DE-SC0025942, and the Gordon and Betty Moore Foundation, grant DOI 10.37807/gbmf12248.
The work of D.A. has been supported by the Scientific and Technological Research
Council of Türkiye (TÜBİTAK) 2232-B International Fellowship for Early Stage Researchers
Programme grant number 122C341, and by the COST Action within the project COSMIC WISPers (Grant No. CA21106). 
The work of D.B. and A.W. has been supported by the Cluster of Excellence “Precision Physics,
Fundamental Interactions, and Structure of Matter” (PRISMA++ EXC 2118/2) funded
by the German Research Foundation (DFG) within the German Excellence Strategy
(Project ID 390831469) and by the COST Action within the project COSMIC WISPers (Grant No. CA21106).
DFJK acknowledges support from the U.S. National Science Foundation under grant PHYS-2510625.

\clearpage

\appendix

\section{Absoprtion of microwave radiation by the electron spin ensemble}\label{app:A}

\noindent
The EPR experiment probes the interaction between the electron spin ensemble and the applied bias magnetic field $B$~\cite{Abragam1970}. This is done by measuring the absorption of resonant microwave radiation by the spin sample placed inside a cavity, fig.~1 of main text. 
Let us start by considering the dynamics of a single electron spin $s=1/2$ inside such a sample.
We use Fermi's golden rule to find the transition rate $w_{01}$ between the spin-down state $\ket{0}$ and spin-up state $\ket{1}$:
\begin{equation}
    w_{01}=\frac{2 \pi}{\hbar} |H_{01}|^2 \rho
    \label{eq:fgr1}
\end{equation}
where $H_{01} $ is the transition matrix element and $\rho$ is the usual density of states factor.

The transitions are driven by the spin-1/2 Zeeman interaction matrix element $H_{01} = \hbar\gamma B_1/2$, where $\gamma$ is the electron spin gyromagnetic ratio and $B_1$ is the amplitude of the EPR microwave field. We make the rotating wave approximation, which keeps only the component $B_1/2$ of the linearly-polarized microwave field, co-rotating with the spin. We express the density of states as a function of angular frequency $\omega = 2 \pi \nu$ of the microwave radiation: 
\begin{equation}
\rho=h(\omega-\omega')/\hbar,
\label{eq:dos}
\end{equation}
where $h(\omega-\omega')$ is the single-spin (homogeneous) absorption spectrum and $\omega'$ is the spin transition angular frequency~\cite{Abragam1961}. 
This gives the transition rate
\begin{equation}
    w_{01}=\frac{\pi\gamma^2B_1^2}{8} h(\omega-\omega')
    \label{eq:fgr2}
\end{equation}

The microwave power absorbed by the spin is given by
%\footnote{Abragam-Bleaney 2.52a}
\begin{equation}
\frac{dW_1}{dt}=(a_0-a_1)w_{01}\hbar \omega 
\label{eq:power}
\end{equation}
where $a_0$ and $a_1$ are the populations of the spin down and up states.
The Boltzmann distribution at temperature $T$ gives:
\begin{equation}
    a_1-a_0 \approx 1-e^{-\hbar \omega/k_B T} \approx \frac{\hbar \omega}{k_B T}, \label{eq:Boltzmann}
\end{equation}
assuming $\omega'\approx\omega$, $\hbar\omega\ll k_BT$, which is true for our measurements.
Combining equations (\ref{eq:fgr2}), (\ref{eq:power}) and (\ref{eq:Boltzmann}) gives:
\begin{equation}
    \frac{dW_1}{dt}=
    \frac{\pi \hbar^2\omega^2 \gamma^2 B_1^2}{8k_BT} h(\omega-\omega').
    \label{eq:power2}
\end{equation}

Let us consider the entire spin ensemble. 
If the absorption lineshape is dominated by the homogeneous broadening, the absorbed power is given by eq.~(\ref{eq:power2}). However, in general, inhomogeneous broadening can result from shifts of transition frequencies of different spins in the ensemble. Let us define the probability density $f(\omega')$ of having the spin resonance occur at frequency $\omega'$. This is normalized such that 
\begin{equation}
\int_{-\infty}^{\infty} f(\omega') d \omega'=1.
\label{eq:Q}
\end{equation}
Then the inhomogeneously-broadened absorption spectrum is given by
\begin{equation}
L(\omega) = \int_{-\infty}^{\infty} f(\omega') h(\omega-\omega') d \omega'.
\label{eq:conv}
\end{equation}
Therefore the power absorbed by the entire ensemble of $N$ electron spins becomes:
\begin{equation}
    \frac{dW}{dt}=
    \frac{\pi \hbar^2\omega^2 \gamma^2 B_1^2}{8k_BT} N L(\omega),
    \label{eq:power3}
\end{equation}

In order to connect to experimental measurements, it is convenient to express the absorbed power in terms of the (dimensionless) imaginary part of the sample's magnetic susceptibility $\chi''$:
%\cite{Abragam1970} \footnote{A-B: 2.59}
\begin{equation}
\frac{dW}{dt}=\frac{V}{2}\omega\chi'' B_1^2    
\label{eq:pandsusc}
\end{equation}
where $V$ is the sample volume. 
%The $1/2$ factor accounts for linear polarization of the incident microwave field.
Therefore
\begin{equation}
    \chi''(\omega)=\frac{\pi \hbar^2\gamma^2 n\omega}{4k_BT}L(\omega),
    \label{eqn:susc}
\end{equation}
where $n=N/V$ is the electron spin number density.

\section{Spin ensemble absorption lineshapes}
\label{app:B}

\subsection{The homogeneous lineshape} \label{sec:2A}

\noindent
The homogeneous spin absorption lineshape is given by the Lorenzian function:
\begin{equation}
h(\omega-\omega') = \frac{1}{\pi} \frac{\tau_2}{1 + \gamma^ 2 B_1^2 \tau_1 \tau_2 + (\omega-\omega')^2\tau_2^2},
\label{eqn:Lor}
\end{equation}
where $\omega'$ is the spin resonance angular frequency, $\tau_1$ is the spin population relaxation time, and $\tau_2$ is the spin coherence relaxation time.
The term $\gamma^2 B_1^2 \tau_1 \tau_2$ is responsible for spin saturation by the microwave drive. The lineshape function is normalized such that
\begin{equation}
\int_{-\infty}^{\infty}h(\omega)\,d\omega = \frac{1}{\sqrt{1 + \gamma ^ 2 B_1^2 \tau_1 \tau_2}}.
\label{eqn:LorNorm}
\end{equation}

\subsection{The overall lineshape}

The overall absorption lineshape $L(\omega)$ is given by the convolution (\ref{eq:conv}) of the inhomogeneous $f(\omega')$ with the homogeneous lineshape $h(\omega-\omega')$, given by eq.~(\ref{eqn:Lor}). 
We estimate the width of the homogeneous line from the experimental results obtained with PbTiO$_3$ sample. The EPR absorption linewidths measured for PMN-PT are significantly broader, therefore, for PMN-PT, the inhomogeneous broadening dominates. Thus we approximate the homogeneous lineshape as a delta-function~\cite{Portis1953}:
\begin{align}
    h(\omega - \omega') \approx \frac{\delta(\omega-\omega')}{\sqrt{1+\gamma^2 B_1^2 \tau_1 \tau_2}}.
\end{align}
The overall absorption lineshape becomes:
\begin{equation}
L(\omega) =\int_0 ^\infty f(\omega')  \frac{\delta(\omega-\omega')}{\sqrt{1+\gamma^2 B_1^2 \tau_1 \tau_2}} d \omega' =  \frac{f(\omega)}{\sqrt{1 + \gamma^2 B_1 ^ 2 \tau_1 \tau_2}}.
\label{eq:inhbr}
\end{equation}

In our EPR experiments the microwave frequency is fixed, and the bias magnetic field $B$ is varied. Thus it is convenient to work in magnetic field units, expressing frequencies in terms of corresponding magnetic field: $B = \omega/\gamma$. Then
\begin{equation}
L(B) =\frac{f(B)}{\sqrt{1 + \gamma^2 B_1 ^ 2 \tau_1 \tau_2}}.
\label{eq:inhbr}
\end{equation}

\subsection{Inhomogeneous broadening due to g-factor anisotropy}

\noindent
The dominant static term in the electron spin Hamiltonian is the Zeeman interaction:
\begin{equation}
H_Z = \mu_B \sum_{\alpha\beta}g_{\alpha\beta}B_{\alpha}S_{\beta}
\end{equation}
where $\mu_B$ is the Bohr magneton, $B$ is the bias magnetic field, $g$ is the g-tensor, $S$ is the electronic spin operator, and the sum is over Cartesian components $\alpha,\beta\in \{x,y,z\}$.

The electronic g-tensor in the materials we study can be accurately approximated to be uniaxial~\cite{Warren1993}. We denote its principal values as $g_\perp$ (perpendicular to symmetry axis) and $g_\parallel$ (parallel to symmetry axis)~\cite{Abragam1970}. We choose the coordinate system with the z-axis that points along the bias magnetic field. If, for a particular spin, the crystal symmetry axis makes an angle $\theta$ with the magnetic field, the static Zeeman Hamiltonian for this spin is given by
\begin{equation}
H_Z = \mu_B\sqrt{g_\parallel^2 \cos^2\theta + g_\bot^2 \sin^2\theta } BS_z,
\label{eqn:g2D}
\end{equation}
and the spin transition angular frequency is given by
\begin{equation}
\hbar \omega' = \mu_B B \sqrt{g_\parallel^2 \cos^2\theta + g_\bot^2 \sin^2\theta}
\label{eqn:Hamtrans}
\end{equation}

In our work we study crystalline samples of PbTiO$_3$ and PMN-PT. Nevertheless, there is an inevitable disorder in the samples, caused, for example by the ferroelectric domain structure. This is especially true for the relaxor ferroelectric PMN-PT~\cite{Wang2014a} \ja{[fix ref]}. Therefore we make the assumption that the spin symmetry axes are randomly distributed throughout the sample. Let us calculate the probability distribution $f_p(\omega')$ that describes the probability of a spin undergoing resonance at a frequency $\omega'$. We start with the number of spins $dN'$ with their symmetry axis between $\theta$ and $\theta + d\theta$:
\begin{equation}
    dN' = N\sin \theta d\theta
    \label{eqn:dN2D}
\end{equation}
where $N$ is the number of all spins in the sample. 
Thus the probability density of funding a spin with its symmetry axis at angle $\theta$ is
\begin{equation}
    f_p(\theta) = \frac{1}{N}\frac{dN'}{d \theta} = \sin \theta
    \label{eqn:p(th)}
\end{equation}

\begin{figure}[h!]
  \centering
 \includegraphics[width=0.7\linewidth]{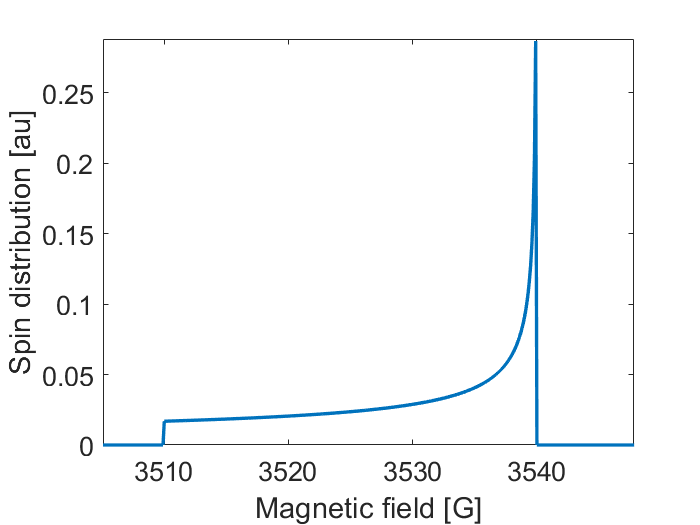}    
\caption{The power lineshape with $B_{\bot} = 3510$ G and $B_{\parallel} = 3540$ G.}
\label{fig:dNdB}
\end{figure}

In our EPR experiments the microwave frequency is usually fixed and the bias magnetic field $B$ is varied. Thus it is convenient to work in magnetic field units, expressing frequencies in terms of corresponding magnetic field.
Let us introduce
\begin{align}
    B_\bot &= \frac{\hbar \omega'}{\mu_B g_\bot},\\ B_\parallel &= \frac{\hbar \omega'}{\mu_B g_\parallel}.
    \label{eqn:p(th)}
\end{align}
We can now re-write eq.~(\ref{eqn:Hamtrans}) as:
\begin{equation}
    \cos{\theta} = \frac{B_\parallel}{\sqrt{B_\bot ^ 2 - B_\parallel ^2}} \frac{\sqrt{B_\bot ^ 2 - B^2}}{B}.
\end{equation}
This equation gives the angle $\theta$ between the applied field and the crystal axis for spins that are resonant at field $B$. 
Differentiating both sides of the equation and rearranging gives:
\begin{equation} 
    \sin{\theta} \, d\theta = \frac{B_\parallel B_\bot^2}{B^2 \sqrt{(B_\parallel ^2 - B_\bot ^ 2) (B ^ 2 - B_\bot ^ 2)}}    dB
\end{equation}
Inserting this equation into eq.~(\ref{eqn:dN2D}), we get the expression for the probability density as a function of resonance field $B$:
\begin{equation} 
    f_p(B) \equiv \frac{1}{N}\frac{dN'}{dB} = \frac{1}{B^2}\frac{B_\parallel B_\bot ^2}{\sqrt{(B_\parallel^2-B_\bot^2)(B^2 - B_\bot^2)}}
    \label{eqn:2ddNdB}
\end{equation}
The function is only defined in the range $ B_\bot < B < B_\parallel $ or $ B_\bot > B > B_\parallel $.
It is normalized such that the integral over $B$ is $1$.
This is the powder lineshape that describes the inhomogeneous distribution of spin resonance frequencies in an ensemble of randomly-oriented crystallites, fig.~\ref{fig:dNdB}.

\subsection{Inhomogeneous broadening due to unresolved isotropic hyperfine interaction}

The PMN-PT crystal lattice has a high density of nuclear spins, dominated by $^{93}$Nb ($I=9/2$, 100\% abundance, gyromagnetic ratio $2\pi\times 10\uu{MHz/T}$, nuclear quadrupole moment -0.32\uu{barn}).  Due to the large number of nuclear spins that interact with any given paramagnetic center, the individual hyperfine transitions can not be resolved and they coalesce into a broad envelope. We also note that the nuclear spin Hamiltonian contains a large quadrupolar interaction term, along the crystal symmetry axis.

We approximate the inhomogeneous broadening of the EPR transition due to the unresolved hyperfine structure by the Gaussian function~\cite{Bales1998}:
\begin{equation}
f_{hf}(\omega') = f_{hf}(\omega'-\omega_0;\Gamma) = \frac{1}{\sqrt{2\pi} \Gamma} \exp\Bigg [-\frac{(\omega'-\omega_0)^2}{2 \Gamma ^ 2 } \Bigg ],
\label{eq:Gau}
\end{equation}
where $\Gamma$ is the linewidth of the unresolved hyperfine broadening, and $\omega_0$ is the mean value of the EPR transition angular frequency at a given bias magnetic field. As in the previous section, in an EPR experiment the microwave frequency is fixed and the bias magnetic field is varied. Therefore it is convenient to express the lineshape in terms of the magnetic field:
\begin{equation}
f_{hf}(B') = f_{hf}(B'-B_0;\Delta) = \frac{1}{\sqrt{2\pi} \Delta} \exp\Bigg [-\frac{(B'-B_0)^2}{2 \Delta ^ 2 } \Bigg ],
\label{eq:GauB}
\end{equation}
where $\Delta = \Gamma/\gamma$ is the linewidth due to unresolved hyperfine broadening expressed in magnetic field units and $B_0=\omega_0/\gamma$.

\subsection{Overall lineshapes in PMN-PT EPR}

In this study we model two populations of light-induced paramagnetic centers: localized on the Pb$^{3+}$ and on the Ti$^{3+}$ ions.
The electron spins localized on the Pb$^{3+}$ ions have isotropic g-factors, because the electrons occupy symmetric s-orbitals~\cite{Warren1993}. The dominant source of inhomogeneous broadening is the unresolved hyperfine interaction with neighboring nuclear spins, with the corresponding lineshape $f_{hf}(\omega')$ modeled with eq.~(\ref{eq:GauB}). Therefore in our work the overall absorption lineshape for these spins is modeled as
\begin{equation}
L_{Pb}(B) =\frac{1}{\sqrt{1 + \gamma^2 B_1 ^ 2 \tau_{1P} \tau_{2P}}}\frac{1}{\sqrt{2\pi} \Delta_{Pb}} \exp\Bigg [-\frac{(B-B_{Pb})^2}{2 \Delta_{Pb} ^ 2 } \Bigg ],
\label{eq:PbGB}
\end{equation}
where $B_{Pb}$ is the line center, $\Delta_{Pb}$ is the linewidth, and $\tau_{1P},\,\tau_{2P}$ are the relaxation parameters defined in section \ref{sec:2A}.

The electron spins localized on the Ti$^{3+}$ ions occupy d-orbitals, which gives rise to g-factor anisotropy~\cite{Warren1992}. Their overall absorption lineshape is the convolution of the hyperfine-broadened lineshape and the powder lineshape:
\begin{align}
    L_{Ti}(B) &=  \frac{1}{\sqrt{1 + \gamma ^ 2 B_1^2 \tau_{1T} \tau_{2T}}} \int f_{hf}(B-B';\Delta_{T})  f_p(B') dB' \nonumber \\ 
    &= \frac{1}{\sqrt{1 + \gamma ^ 2 B_1^2 \tau_{1T} \tau_{2T}}} \int \frac{f_p(B')}{\sqrt{2\pi} \Delta_{Ti}} \exp\Bigg [-\frac{(B-B')^2}{2 \Delta_{Ti} ^ 2 } \Bigg ] dB'.
    \label{eq:TiGB}
\end{align}

We model the measured EPR signals as the sum of the signals due to paramagnetic centers localized on the Pb$^{3+}$ and the Ti$^{3+}$ ions.

\section{Extracting parameters from EPR spectra}
\label{app:C}

\subsection{The microwave power dependence}

In this section, we consider how the EPR spectrum depends on the microwave power.
This will allow us to use our data to extract the spin densities and the product of $\tau_1 \tau_2$, and therefore estimate the contribution of electrons to the nuclear $T_1$ time in separate NMR experiments.

The voltage detected by an EPR spectrometer is~\cite{Eaton2010}
\begin{equation}
    V_s= \alpha \frac{d\chi''}{dB} \eta Q \sqrt{P Z_0},
    \label{eq:voltage}
\end{equation}
where $\alpha$ is the modulation amplitude of the bias magnetic field $B$, $\eta$ is the filling factor, $Q$ is the loaded quality factor of the resonator, $P$ is the microwave power, and $Z_0$ is the characteristic impedance of the transmission line. In EPR spectroscopy, in order to increase the SNR, the static magnetic field is modulated with a modulation amplitude $\alpha$ and phase-sensitive detection is used. 
As a consequence, the observed signal is the derivative of the spectrum, $\frac{d \chi''}{d B }$ and the signal is proportional to the modulation amplitude, as long as the modulation amplitude is less than the signal linewidth (which is true in all our measurements). 

We make use of Eq.~\eqref{eqn:susc} to re-write this voltage as: 
\begin{equation}
    V_s=\frac{\pi n \omega \gamma \hbar^2}{ 4 k_B T} \frac{dL(B)}{dB}\, \eta \, Q \, \alpha \sqrt{P \, Z_0},
    \label{eq:voltage_3}
\end{equation}
where we expressed the lineshape function $L$ in terms of the magnetic field variable by making use of the relation $L(\omega) = L(B)/\gamma$, which follows from normalization.

Let us separate the voltage contributions from the electron spins localized on the Pb$^{3+}$ and on the Ti$^{3+}$ ions: 
\begin{equation}
    V_s=V_{Pb} + V_{Ti} = \frac{\pi \omega \gamma \hbar^2}{ 4 k_B T} \, \eta \, Q \, \alpha \sqrt{P \, Z_0} \Big( n_{Pb} \frac{dL_{Pb}(B)}{dB} + n_{Ti} \frac{dL_{Ti}(B)}{dB} \Big),
    \label{eq:34}
\end{equation}
where lineshapes $L_P$ and $L_T$ are given in Eqs.~\eqref{eq:PbGB},\eqref{eq:TiGB}.

\begin{figure}[h!]
  \centering
 \includegraphics[width=0.7\linewidth]{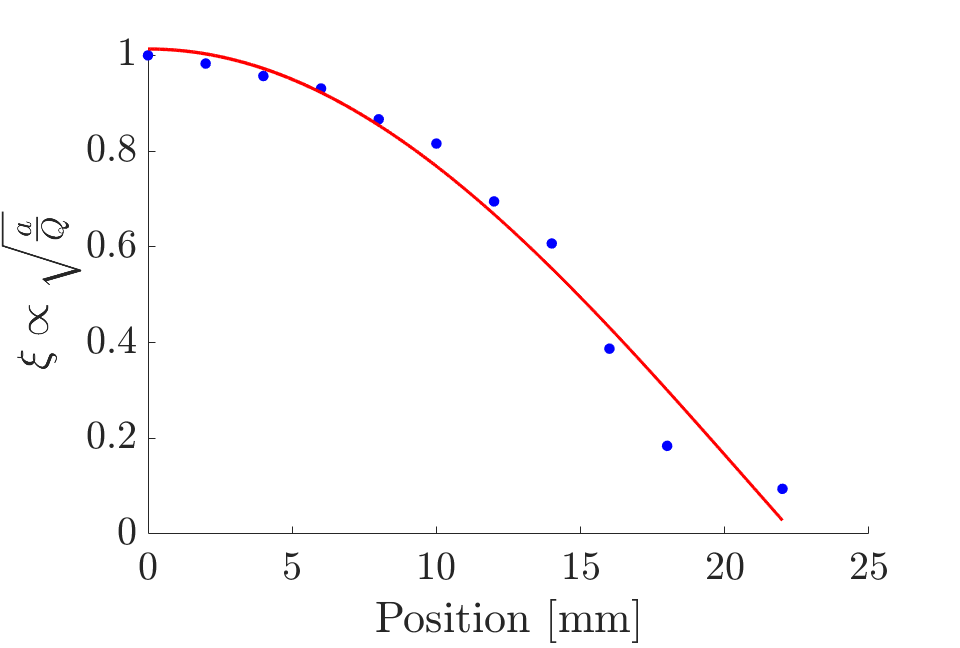}    
\caption{The spatial dependence of the microwave field along the cavity axis. The data points show $\xi \propto \sqrt{a/Q}$ for various values of the sample position $x$, measured from the cavity center. The data are normalized such that $\xi(0)=1$. The line shows a heuristic fit $\xi \propto \cos{(\pi x/2L)}$, where $L = (21 \pm 1)\,$mm is the fit parameter that corresponds to the cavity length.}
\label{fig:cavpos}
\end{figure}

The dependence of the lineshape on the microwave power $P$ is due to the microwave magnetic field amplitude $B_1$: 
\begin{equation}
     B_1  = \Lambda \sqrt{Q P} \xi(x)
    \label{eqn:Lambda}
\end{equation}
where $\xi(x)$ is the function that describes the spatial dependence of $B_1$ on the coordinate $x$ along the vertical axis of the microwave cavity and 
$\Lambda$ is the conversion factor that connects the power injected into the cavity to the resulting magnetic field amplitude~\cite{More1984}. 
The microwave cavity used in our experiments is the Bruker ER 4122 SHQ Super High Q Resonator (TEM$_{011}$ mode). Following the procedure in Ref.~\cite{More1984}, we performed calibration  measurements, described in Sec.~\ref{sec:Cal}, from which we extracted $\Lambda = 0.027\uu{G/\sqrt{W}}$ and $\xi(x)$, shown in Fig.~\ref{fig:cavpos}.

Consider the filling factor, defined as
\begin{equation}
    \eta = \frac{\int_{sample} B_1^2 dV}{\int_{cav} B_1^2 dV}
\end{equation}
where $\int_{sample}$ and $\int_{cav}$ refers to integrals over the sample and the cavity \cite{Eaton2010}. The filling factor quantifies the ratio of the magnetic field energy inside the sample, compared to the whole cavity.
If we assume that the sample dimensions are much smaller than the cavity mode volume, the filling factor takes a simpler form \cite{Dalal1981}:
\begin{equation}
\begin{split}
    \eta = \frac{B_1(x_s)^2 V_s}{\int_{cav} B_1^2 \, dV} = \frac{\Lambda^2 \, P \, Q \, \xi(x_s)^2 \, V_s}{\Lambda^2 P Q \int_{cav} \, \xi(x)^2 \, dV} = \\ \frac{\xi(x_s)^2 \,  V_s}{\int_{cav} \, \xi^2(x) dV } = \frac{\xi(x_s)^2 \, V_s }{V_0}
\end{split}
\end{equation}
where $V_0 = \int_{cav} \xi^2(x) dV $ is the effective volume of the cavity, $V_s$ is the sample volume and $x_s$ is the position of the sample along the vertical axis.

Let us substitute this equation into Eq.~\eqref{eq:34}:
\begin{equation}
    V_s=V_{Pb} + V_{Ti} = \frac{\pi \omega \gamma \hbar^2}{ 4 k_B T} \, \frac{Q\alpha\xi^2(x_s)}{V_0} \sqrt{P \, Z_0} \Big( N_{Pb} \frac{L_{Pb}(B)}{dB} + N_{Ti} \frac{L_{Ti}(B)}{dB} \Big).
    \label{eq:38}
\end{equation}

Let us re-write Eq.~\eqref{eq:38} so that the factors that depend on microwave power are explicitly shown: 
\begin{align}
    V_s=V_{Pb} + V_{Ti} &= \frac{\pi \omega \gamma \hbar^2}{ 4 k_B T} \, \frac{Q\alpha\xi^2(x_s)}{V_0} \sqrt{P \, Z_0} \Bigg( \frac{N_{Pb}}{\sqrt{1 + \gamma^2 B_1^2 \tau_{1,Pb} \tau_{2,Pb}} }\frac{dF_P(B)}{dB} + \frac{N_{Ti}}{\sqrt{1 + \gamma^2 B_1^2 \tau_{1,Ti} \tau_{2,Ti}} }\frac{dF_T(B)}{dB}\Bigg )\nonumber \\
    &= A_{Pb}\frac{d F_{Pb}(B)}{dB} + A_{Ti}\frac{dF_{Ti}(B)}{dB},
     \label{eq:39}
\end{align}
where the coefficients are
\begin{align}
    A_i &= \frac{a_i\sqrt{P}}{\sqrt{1+b_i P}}, \label{eq:AAfit}\\
    a_i &= \frac{\pi \omega \gamma \hbar^2}{ 4 k_B T} \, \frac{Q\alpha\xi^2(x_s)\sqrt{Z_0}}{V_0} N_i, \label{eq:afit} \\  
    b_i &= \gamma^2 \Lambda^2 Q P \xi^2(x_s) \tau_{1,i} \tau_{2,i},
    \label{eq:bfit}
\end{align}
with $i\in\{{\rm Pb,\,Ti}\}$. 
The lineshape function for the Pb$^{3+}$ centers is:
\begin{equation}
F_{Pb}(B) =\frac{1}{\sqrt{2\pi} \Delta_P} \exp\Bigg [-\frac{(B-B_P)^2}{2 \Delta_P ^ 2 } \Bigg ],
\label{eq:PbGB2}
\end{equation}
The lineshape function for the Ti$^{3+}$ centers is:
\begin{align}
    F_{Ti}(B) =  \int \frac{1}{B'^2}\frac{B_\parallel B_\bot ^2}{\sqrt{(B_\parallel^2-B_\bot^2)(B'^2 - B_\bot^2)}}\frac{1}{\sqrt{2\pi} \Delta_T} \exp\Bigg [-\frac{(B-B')^2}{2 \Delta_T ^ 2 } \Bigg ] dB',
    \label{eq:TiGB2}
\end{align}

\subsection{Extracting sample spin densities and relaxation times}

We modeled the measured EPR spectra using Eq.~\eqref{eq:39}. The fit parameters were the weights $A_{Pb},A_{Ti}$, the Pb$^{3+}$ linewidth $\Delta_{Pb}$, and line center $B_{Pb}$ (which we expressed in terms of the g-factor $g_{Pb}=\hbar\omega_0/\mu_BB_{Pb}$), the Ti$^{3+}$ linewidth $\Delta_{Ti}$, and Ti anisotropy parameters $B_\parallel, B_\bot$ (which we expressed in terms of the g-factors $g_{\parallel}=\hbar\omega_0/\mu_BB_\parallel$ and $g_{\bot}=\hbar\omega_0/\mu_BB_\bot$).

The data acquisition and analysis proceeded as follows.
\begin{itemize}
    \item Record EPR spectra at several microwave powers at 10K temperature.
    \item Globally fit all the spectral data sets with a single set of parameters, using Eq.~\eqref{eq:39}, which is the same as Eq. (5) in the main text. The fit parameters are: 2 linewidths ($\Delta_i$), 3 g-factors ($g_{Pb}$ for Pb and $g_\bot , g_\parallel$ for Ti) and amplitudes $A_{Pb},A_{Ti}$ at all powers. We assume that g-factors and linewidths are independent of power.
    \item The amplitudes $A_i$ were fitted with eq.~\eqref{eq:AAfit},
    which allows us to extract $a_i, b_i$ parameters.
\end{itemize}

The product of relaxation times $\tau_{1,i} \tau_{2,i}$ was extracted from the best-fit parameter $b_i$ using Eq. \eqref{eq:bfit}. If $\tau_{2,i}$ is known or estimated independently, $\tau_{1,i}$ can be determined.

The number of spins $N_i$ could be extracted from the best-fit parameter $a_i$ using Eq. \eqref{eq:afit}. However this would require accurate knowledge of the other parameters in this equation. Instead, we recorded the same EPR data for a calibration sample, with a known electron spin density, and compared it with the EPR data for the sample of interest. The spin number in the sample of interest was then extracted from this comparison.

As the calibration sample we used the solution of the free radical TEMPO. This radical was first dissolved in acetone at known concentration and then dried out. This calibration sample contains $N_r=4 \times 10^{16}$ spins.
We recorded the EPR spectra of the calibration sample at 10K and at different microwave powers, using the procedure described above. After fitting the spectra, we obtain the best-fit parameter $a_r$ and extract the spin numbers in the PMN-PT sampe using
\begin{align}
    N_{i} = N_{r}\frac{a_{i}}{a_{r}}\frac{\xi^2(x_r)}{\xi^2(x_i)}\frac{Q_{r}}{Q_{i}}\frac{\alpha_{r}}{\alpha_{i}}
    \label{eq:apera}
\end{align}
where subscript $r$ indicates radicals and $i$ indicates the Pb$^{3+}$ or the Ti$^{3+}$ spin populations.
The radical sample was placed at the center of the cavity, where the magnetic field is maximum, therefore $\xi(x_r)=1$.

The uncertainties in spin density were calculated using the 
error propagataion formula:
\begin{align}
    \frac{\sigma_N^2}{N^2} = \frac{\sigma_{Nr}^2}{N_r^2} + \frac{\sigma_a^2}{a^2} + \frac{\sigma_{a_r}^2}{a_r^2} +
    4 \frac{\sigma_\xi^2}{\xi^2} +
    \frac{\sigma_Q^2}{Q^2} + \frac{\sigma_{Q_r}^2}{Q_r^2}
\end{align}
The total systematic uncertainty is $\frac{\sigma_N}{N} = 0.23$, with the following contributions.
\begin{itemize}
    \item $\frac{\sigma_{N_r}}{N_r} = 0.1$, the calibration sample is a free radical with a 10\% uncertainty in the number of spins.
    \item $\frac{\sigma_a}{a}$ and $\frac{\sigma_{a_r}}{a_r}$ are small 0.008, 0.002.
    \item $2 \frac{\sigma_\xi}{\xi } = 0.12$, due to 1 mm uncertainty in sample position, see Fig.~\ref{fig:cavpos}. 
    \item $\frac{\sigma_Q}{Q} = 0.13$, given by the change in cavity $Q$ due to 1 mm uncertainty in sample position.
    \item $\frac{\sigma_{Q_r}}{Q_r}= 0.1$, same as in the previous point.
\end{itemize}

\subsection{Calibration of $\Lambda$ and $\xi(x)$} \label{sec:Cal}

Dielectric losses in the PMT-PT sample prevented us from being able to tune the cavity if the sample was placed at the cavity center. Therefore the sample was placed $15\uu{mm}$ away from the center of the cavity, where the microwave field was smaller and the cavity loss was tolerable. In order to estimate the number of spins and relaxation times it was necessary to calibrate the magnetic field distribution $\xi(x)$ inside the cavity. 
Equation \eqref{eq:afit} shows that $a \propto Q \xi ^ 2$, therefore measuring $a$ and $Q$ allows us to calibrate $\xi$. 
We performed this measurements using the same TEMPO radical sample that we used for spin number calibration. EPR spectra were recorded with this sample positioned at various locations along the axis of the microwave cavity. For each spectrum we recorded the cavity quality factor $Q$ and the best-fit amplitude $a$. The resulting $\xi(x)$ is shown in Fig.~\ref{fig:cavpos}, with the normalization $\xi(0)=1$ at the center of the cavity. From this plot we extract $\xi(15\uu{mm})=0.50 \pm 0.06$, assuming $\pm 0.5\uu{mm}$ uncertainty in PMN-PT sample position.

The conversion factor $\Lambda$ converts the microwave power $P$ to the magnetic field amplitude $B_1$, Eq.~\eqref{eqn:Lambda}.
In order to calibrate $\Lambda$ we performed EPR experiments on the sample of Fremy's salt with known values of $\tau_1 = 0.33\uu{\mu s}$ and $\tau_2 = 0.25\uu{\mu s}$~\cite{More1984,Schreurs1961}. The Fremy's salt sample was prepared with $6 \times 10^{17} cm^{-3}$ spins in a 0.015 ml volume and concentration $M = 10^{-3}$ mol/L. We acquired EPR spectra for a range of microwave powers, including power values where saturation was observed. We fit the spectra and extracted the best-fit parameter $b$. We then used Eq.~\eqref{eq:bfit} and the quoted values of $\tau_1$ and $\tau_2$ to extract the value $\Lambda = 0.0273\uu{G/\sqrt{W}}$.

\bibliography{library}

\begin{thebibliography}{44}%
\makeatletter
\providecommand \@ifxundefined [1]{%
 \@ifx{#1\undefined}
}%
\providecommand \@ifnum [1]{%
 \ifnum #1\expandafter \@firstoftwo
 \else \expandafter \@secondoftwo
 \fi
}%
\providecommand \@ifx [1]{%
 \ifx #1\expandafter \@firstoftwo
 \else \expandafter \@secondoftwo
 \fi
}%
\providecommand \natexlab [1]{#1}%
\providecommand \enquote  [1]{``#1''}%
\providecommand \bibnamefont  [1]{#1}%
\providecommand \bibfnamefont [1]{#1}%
\providecommand \citenamefont [1]{#1}%
\providecommand \href@noop [0]{\@secondoftwo}%
\providecommand \href [0]{\begingroup \@sanitize@url \@href}%
\providecommand \@href[1]{\@@startlink{#1}\@@href}%
\providecommand \@@href[1]{\endgroup#1\@@endlink}%
\providecommand \@sanitize@url [0]{\catcode `\\12\catcode `\$12\catcode
  `\&12\catcode `\#12\catcode `\^12\catcode `\_12\catcode `\%12\relax}%
\providecommand \@@startlink[1]{}%
\providecommand \@@endlink[0]{}%
\providecommand \url  [0]{\begingroup\@sanitize@url \@url }%
\providecommand \@url [1]{\endgroup\@href {#1}{\urlprefix }}%
\providecommand \urlprefix  [0]{URL }%
\providecommand \Eprint [0]{\href }%
\providecommand \doibase [0]{https://doi.org/}%
\providecommand \selectlanguage [0]{\@gobble}%
\providecommand \bibinfo  [0]{\@secondoftwo}%
\providecommand \bibfield  [0]{\@secondoftwo}%
\providecommand \translation [1]{[#1]}%
\providecommand \BibitemOpen [0]{}%
\providecommand \bibitemStop [0]{}%
\providecommand \bibitemNoStop [0]{.\EOS\space}%
\providecommand \EOS [0]{\spacefactor3000\relax}%
\providecommand \BibitemShut  [1]{\csname bibitem#1\endcsname}%
\let\auto@bib@innerbib\@empty
%</preamble>
\bibitem [{\citenamefont {Degen}\ \emph {et~al.}(2017)\citenamefont {Degen},
  \citenamefont {Reinhard},\ and\ \citenamefont {Cappellaro}}]{Degen2017}%
  \BibitemOpen
  \bibfield  {author} {\bibinfo {author} {\bibfnamefont {C.~L.}\ \bibnamefont
  {Degen}}, \bibinfo {author} {\bibfnamefont {F.}~\bibnamefont {Reinhard}},\
  and\ \bibinfo {author} {\bibfnamefont {P.}~\bibnamefont {Cappellaro}},\
  }\href {https://doi.org/10.1103/RevModPhys.89.035002} {\bibfield  {journal}
  {\bibinfo  {journal} {Reviews of Modern Physics}\ }\textbf {\bibinfo {volume}
  {89}},\ \bibinfo {pages} {035002} (\bibinfo {year} {2017})},\ \Eprint
  {https://arxiv.org/abs/1611.02427} {arXiv:1611.02427} \BibitemShut {NoStop}%
\bibitem [{\citenamefont {Safronova}\ \emph {et~al.}(2018)\citenamefont
  {Safronova}, \citenamefont {Budker}, \citenamefont {Demille}, \citenamefont
  {Kimball}, \citenamefont {Derevianko},\ and\ \citenamefont
  {Clark}}]{Safronova2018}%
  \BibitemOpen
  \bibfield  {author} {\bibinfo {author} {\bibfnamefont {M.~S.}\ \bibnamefont
  {Safronova}}, \bibinfo {author} {\bibfnamefont {D.}~\bibnamefont {Budker}},
  \bibinfo {author} {\bibfnamefont {D.}~\bibnamefont {Demille}}, \bibinfo
  {author} {\bibfnamefont {D.~F.}\ \bibnamefont {Kimball}}, \bibinfo {author}
  {\bibfnamefont {A.}~\bibnamefont {Derevianko}},\ and\ \bibinfo {author}
  {\bibfnamefont {C.~W.}\ \bibnamefont {Clark}},\ }\href
  {https://doi.org/10.1103/RevModPhys.90.025008} {\bibfield  {journal}
  {\bibinfo  {journal} {Reviews of Modern Physics}\ }\textbf {\bibinfo {volume}
  {90}},\ \bibinfo {pages} {025008} (\bibinfo {year} {2018})},\ \Eprint
  {https://arxiv.org/abs/1710.01833} {arXiv:1710.01833} \BibitemShut {NoStop}%
\bibitem [{\citenamefont {Kuenstner}\ \emph {et~al.}(2026)\citenamefont
  {Kuenstner}, \citenamefont {Smith}, \citenamefont {Winter}, \citenamefont
  {Ozdemir}, \citenamefont {Mari{\'c}}, \citenamefont {Matthews},\ and\
  \citenamefont {Sushkov}}]{Kuenstner2026}%
  \BibitemOpen
  \bibfield  {author} {\bibinfo {author} {\bibfnamefont {S.~E.}\ \bibnamefont
  {Kuenstner}}, \bibinfo {author} {\bibfnamefont {D.~W.}\ \bibnamefont
  {Smith}}, \bibinfo {author} {\bibfnamefont {A.~J.}\ \bibnamefont {Winter}},
  \bibinfo {author} {\bibfnamefont {E.}~\bibnamefont {Ozdemir}}, \bibinfo
  {author} {\bibfnamefont {T.}~\bibnamefont {Mari{\'c}}}, \bibinfo {author}
  {\bibfnamefont {A.}~\bibnamefont {Matthews}},\ and\ \bibinfo {author}
  {\bibfnamefont {A.~O.}\ \bibnamefont {Sushkov}},\ }\href
  {https://doi.org/10.1038/s41567-026-03187-6} {\bibfield  {journal} {\bibinfo
  {journal} {Nature Physics}\ ,\ \bibinfo {pages}
  {https://doi.org/10.1038/s41567}} (\bibinfo {year} {2026})}\BibitemShut
  {NoStop}%
\bibitem [{\citenamefont {Abragam}(1961)}]{Abragam1961}%
  \BibitemOpen
  \bibfield  {author} {\bibinfo {author} {\bibfnamefont {A.}~\bibnamefont
  {Abragam}},\ }\href@noop {} {\emph {\bibinfo {title} {The {{Principles}} of
  {{Nuclear Magnetism}}}}}\ (\bibinfo  {publisher} {Oxford University Press},\
  \bibinfo {year} {1961})\BibitemShut {NoStop}%
\bibitem [{\citenamefont {Bloembergen}\ \emph {et~al.}(1948)\citenamefont
  {Bloembergen}, \citenamefont {Purcell},\ and\ \citenamefont
  {Pound}}]{Bloembergen1948}%
  \BibitemOpen
  \bibfield  {author} {\bibinfo {author} {\bibfnamefont {N.}~\bibnamefont
  {Bloembergen}}, \bibinfo {author} {\bibfnamefont {E.~M.}\ \bibnamefont
  {Purcell}},\ and\ \bibinfo {author} {\bibfnamefont {R.~V.}\ \bibnamefont
  {Pound}},\ }\href {https://doi.org/10.1103/PhysRev.73.679} {\bibfield
  {journal} {\bibinfo  {journal} {Physical Review}\ }\textbf {\bibinfo {volume}
  {73}},\ \bibinfo {pages} {679} (\bibinfo {year} {1948})}\BibitemShut
  {NoStop}%
\bibitem [{\citenamefont {Bienfait}\ \emph {et~al.}(2016)\citenamefont
  {Bienfait}, \citenamefont {Pla}, \citenamefont {Kubo}, \citenamefont {Zhou},
  \citenamefont {Stern}, \citenamefont {Lo}, \citenamefont {Weis},
  \citenamefont {Schenkel}, \citenamefont {Vion}, \citenamefont {Esteve},
  \citenamefont {Morton},\ and\ \citenamefont {Bertet}}]{Bienfait2016}%
  \BibitemOpen
  \bibfield  {author} {\bibinfo {author} {\bibfnamefont {A.}~\bibnamefont
  {Bienfait}}, \bibinfo {author} {\bibfnamefont {J.~J.}\ \bibnamefont {Pla}},
  \bibinfo {author} {\bibfnamefont {Y.}~\bibnamefont {Kubo}}, \bibinfo {author}
  {\bibfnamefont {X.}~\bibnamefont {Zhou}}, \bibinfo {author} {\bibfnamefont
  {M.}~\bibnamefont {Stern}}, \bibinfo {author} {\bibfnamefont {C.~C.}\
  \bibnamefont {Lo}}, \bibinfo {author} {\bibfnamefont {C.~D.}\ \bibnamefont
  {Weis}}, \bibinfo {author} {\bibfnamefont {T.}~\bibnamefont {Schenkel}},
  \bibinfo {author} {\bibfnamefont {D.}~\bibnamefont {Vion}}, \bibinfo {author}
  {\bibfnamefont {D.}~\bibnamefont {Esteve}}, \bibinfo {author} {\bibfnamefont
  {J.~J.}\ \bibnamefont {Morton}},\ and\ \bibinfo {author} {\bibfnamefont
  {P.}~\bibnamefont {Bertet}},\ }\href {https://doi.org/10.1038/nature16944}
  {\bibfield  {journal} {\bibinfo  {journal} {Nature}\ }\textbf {\bibinfo
  {volume} {531}},\ \bibinfo {pages} {74} (\bibinfo {year} {2016})},\ \Eprint
  {https://arxiv.org/abs/1508.06148} {arXiv:1508.06148} \BibitemShut {NoStop}%
\bibitem [{\citenamefont {Ernst}\ \emph {et~al.}(1990)\citenamefont {Ernst},
  \citenamefont {Bodenhausen},\ and\ \citenamefont {Wokaun}}]{Ernst1990a}%
  \BibitemOpen
  \bibfield  {author} {\bibinfo {author} {\bibfnamefont {R.~R.}\ \bibnamefont
  {Ernst}}, \bibinfo {author} {\bibfnamefont {G.}~\bibnamefont {Bodenhausen}},\
  and\ \bibinfo {author} {\bibfnamefont {A.}~\bibnamefont {Wokaun}},\ }\href
  {https://doi.org/10.1093/oso/9780198556473.001.0001} {\emph {\bibinfo {title}
  {Principles of {{Nuclear Magnetic Resonance}} in {{One}} and {{Two
  Dimensions}}}}}\ (\bibinfo  {publisher} {Oxford University Press},\ \bibinfo
  {year} {1990})\BibitemShut {NoStop}%
\bibitem [{\citenamefont {Levitt}(2008)}]{Levitt2008}%
  \BibitemOpen
  \bibfield  {author} {\bibinfo {author} {\bibfnamefont {M.}~\bibnamefont
  {Levitt}},\ }\href@noop {} {\emph {\bibinfo {title} {Spin {{Dynamics}}:
  {{Basics}} of {{Nuclear Magnetic Resonance}}}}}\ (\bibinfo  {publisher}
  {Wiley},\ \bibinfo {year} {2008})\BibitemShut {NoStop}%
\bibitem [{\citenamefont {Albanese}\ \emph {et~al.}(2020)\citenamefont
  {Albanese}, \citenamefont {Probst}, \citenamefont {Ranjan}, \citenamefont
  {Zollitsch}, \citenamefont {Pechal}, \citenamefont {Wallraff}, \citenamefont
  {Morton}, \citenamefont {Vion}, \citenamefont {Esteve}, \citenamefont
  {Flurin},\ and\ \citenamefont {Bertet}}]{Albanese2020}%
  \BibitemOpen
  \bibfield  {author} {\bibinfo {author} {\bibfnamefont {B.}~\bibnamefont
  {Albanese}}, \bibinfo {author} {\bibfnamefont {S.}~\bibnamefont {Probst}},
  \bibinfo {author} {\bibfnamefont {V.}~\bibnamefont {Ranjan}}, \bibinfo
  {author} {\bibfnamefont {C.~W.}\ \bibnamefont {Zollitsch}}, \bibinfo {author}
  {\bibfnamefont {M.}~\bibnamefont {Pechal}}, \bibinfo {author} {\bibfnamefont
  {A.}~\bibnamefont {Wallraff}}, \bibinfo {author} {\bibfnamefont {J.~J.~L.}\
  \bibnamefont {Morton}}, \bibinfo {author} {\bibfnamefont {D.}~\bibnamefont
  {Vion}}, \bibinfo {author} {\bibfnamefont {D.}~\bibnamefont {Esteve}},
  \bibinfo {author} {\bibfnamefont {E.}~\bibnamefont {Flurin}},\ and\ \bibinfo
  {author} {\bibfnamefont {P.}~\bibnamefont {Bertet}},\ }\href
  {https://doi.org/10.1038/s41567-020-0872-2} {\bibfield  {journal} {\bibinfo
  {journal} {Nature Physics}\ }\textbf {\bibinfo {volume} {16}},\ \bibinfo
  {pages} {751} (\bibinfo {year} {2020})}\BibitemShut {NoStop}%
\bibitem [{\citenamefont {Goldman}(1965)}]{Goldman1965}%
  \BibitemOpen
  \bibfield  {author} {\bibinfo {author} {\bibfnamefont {M.}~\bibnamefont
  {Goldman}},\ }\href {https://doi.org/10.1103/PhysRev.138.A1675} {\bibfield
  {journal} {\bibinfo  {journal} {Physical Review}\ }\textbf {\bibinfo {volume}
  {138}},\ \bibinfo {pages} {A1675} (\bibinfo {year} {1965})}\BibitemShut
  {NoStop}%
\bibitem [{\citenamefont {{van Kesteren}}\ \emph {et~al.}(1985)\citenamefont
  {{van Kesteren}}, \citenamefont {Wenckebach},\ and\ \citenamefont
  {Schmidt}}]{VanKesteren1985}%
  \BibitemOpen
  \bibfield  {author} {\bibinfo {author} {\bibfnamefont {H.~W.}\ \bibnamefont
  {{van Kesteren}}}, \bibinfo {author} {\bibfnamefont {W.~{\relax Th}.}\
  \bibnamefont {Wenckebach}},\ and\ \bibinfo {author} {\bibfnamefont
  {J.}~\bibnamefont {Schmidt}},\ }\href
  {https://doi.org/10.1103/PhysRevLett.55.1642} {\bibfield  {journal} {\bibinfo
   {journal} {Physical Review Letters}\ }\textbf {\bibinfo {volume} {55}},\
  \bibinfo {pages} {1642} (\bibinfo {year} {1985})}\BibitemShut {NoStop}%
\bibitem [{\citenamefont {Diller}\ \emph {et~al.}(2007)\citenamefont {Diller},
  \citenamefont {Prakash}, \citenamefont {Alia}, \citenamefont {Gast},
  \citenamefont {Matysik},\ and\ \citenamefont {Jeschke}}]{Diller2007}%
  \BibitemOpen
  \bibfield  {author} {\bibinfo {author} {\bibfnamefont {A.}~\bibnamefont
  {Diller}}, \bibinfo {author} {\bibfnamefont {S.}~\bibnamefont {Prakash}},
  \bibinfo {author} {\bibfnamefont {A.}~\bibnamefont {Alia}}, \bibinfo {author}
  {\bibfnamefont {P.}~\bibnamefont {Gast}}, \bibinfo {author} {\bibfnamefont
  {J.}~\bibnamefont {Matysik}},\ and\ \bibinfo {author} {\bibfnamefont
  {G.}~\bibnamefont {Jeschke}},\ }\href {https://doi.org/10.1021/jp072428r}
  {\bibfield  {journal} {\bibinfo  {journal} {The Journal of Physical Chemistry
  B}\ }\textbf {\bibinfo {volume} {111}},\ \bibinfo {pages} {10606} (\bibinfo
  {year} {2007})}\BibitemShut {NoStop}%
\bibitem [{\citenamefont {Dommaschk}\ \emph {et~al.}(2015)\citenamefont
  {Dommaschk}, \citenamefont {Peters}, \citenamefont {Gutzeit}, \citenamefont
  {Sch{\"u}tt}, \citenamefont {N{\"a}ther}, \citenamefont {S{\"o}nnichsen},
  \citenamefont {Tiwari}, \citenamefont {Riedel}, \citenamefont {Boretius},\
  and\ \citenamefont {Herges}}]{Dommaschk2015}%
  \BibitemOpen
  \bibfield  {author} {\bibinfo {author} {\bibfnamefont {M.}~\bibnamefont
  {Dommaschk}}, \bibinfo {author} {\bibfnamefont {M.}~\bibnamefont {Peters}},
  \bibinfo {author} {\bibfnamefont {F.}~\bibnamefont {Gutzeit}}, \bibinfo
  {author} {\bibfnamefont {C.}~\bibnamefont {Sch{\"u}tt}}, \bibinfo {author}
  {\bibfnamefont {C.}~\bibnamefont {N{\"a}ther}}, \bibinfo {author}
  {\bibfnamefont {F.~D.}\ \bibnamefont {S{\"o}nnichsen}}, \bibinfo {author}
  {\bibfnamefont {S.}~\bibnamefont {Tiwari}}, \bibinfo {author} {\bibfnamefont
  {C.}~\bibnamefont {Riedel}}, \bibinfo {author} {\bibfnamefont
  {S.}~\bibnamefont {Boretius}},\ and\ \bibinfo {author} {\bibfnamefont
  {R.}~\bibnamefont {Herges}},\ }\href {https://doi.org/10.1021/jacs.5b00929}
  {\bibfield  {journal} {\bibinfo  {journal} {Journal of the American Chemical
  Society}\ }\textbf {\bibinfo {volume} {137}},\ \bibinfo {pages} {7552}
  (\bibinfo {year} {2015})}\BibitemShut {NoStop}%
\bibitem [{\citenamefont {Michalak}\ \emph {et~al.}(2011)\citenamefont
  {Michalak}, \citenamefont {Xu}, \citenamefont {Lowery}, \citenamefont
  {Crawford}, \citenamefont {Ledbetter}, \citenamefont {Bouchard},
  \citenamefont {Wemmer}, \citenamefont {Budker},\ and\ \citenamefont
  {Pines}}]{Michalak2011}%
  \BibitemOpen
  \bibfield  {author} {\bibinfo {author} {\bibfnamefont {D.~J.}\ \bibnamefont
  {Michalak}}, \bibinfo {author} {\bibfnamefont {S.}~\bibnamefont {Xu}},
  \bibinfo {author} {\bibfnamefont {T.~J.}\ \bibnamefont {Lowery}}, \bibinfo
  {author} {\bibfnamefont {C.~W.}\ \bibnamefont {Crawford}}, \bibinfo {author}
  {\bibfnamefont {M.}~\bibnamefont {Ledbetter}}, \bibinfo {author}
  {\bibfnamefont {L.-S.}\ \bibnamefont {Bouchard}}, \bibinfo {author}
  {\bibfnamefont {D.~E.}\ \bibnamefont {Wemmer}}, \bibinfo {author}
  {\bibfnamefont {D.}~\bibnamefont {Budker}},\ and\ \bibinfo {author}
  {\bibfnamefont {A.}~\bibnamefont {Pines}},\ }\href
  {https://doi.org/10.1002/mrm.22811} {\bibfield  {journal} {\bibinfo
  {journal} {Magnetic Resonance in Medicine}\ }\textbf {\bibinfo {volume}
  {66}},\ \bibinfo {pages} {603} (\bibinfo {year} {2011})}\BibitemShut
  {NoStop}%
\bibitem [{\citenamefont {Budker}\ \emph {et~al.}(2014)\citenamefont {Budker},
  \citenamefont {Graham}, \citenamefont {Ledbetter}, \citenamefont
  {Rajendran},\ and\ \citenamefont {Sushkov}}]{Budker2014}%
  \BibitemOpen
  \bibfield  {author} {\bibinfo {author} {\bibfnamefont {D.}~\bibnamefont
  {Budker}}, \bibinfo {author} {\bibfnamefont {P.~W.}\ \bibnamefont {Graham}},
  \bibinfo {author} {\bibfnamefont {M.}~\bibnamefont {Ledbetter}}, \bibinfo
  {author} {\bibfnamefont {S.}~\bibnamefont {Rajendran}},\ and\ \bibinfo
  {author} {\bibfnamefont {A.~O.}\ \bibnamefont {Sushkov}},\ }\href
  {https://doi.org/10.1103/PhysRevX.4.021030} {\bibfield  {journal} {\bibinfo
  {journal} {Physical Review X}\ }\textbf {\bibinfo {volume} {4}},\ \bibinfo
  {pages} {021030} (\bibinfo {year} {2014})}\BibitemShut {NoStop}%
\bibitem [{\citenamefont {Aybas}\ \emph
  {et~al.}(2021{\natexlab{a}})\citenamefont {Aybas}, \citenamefont {Adam},
  \citenamefont {Blumenthal}, \citenamefont {Gramolin}, \citenamefont
  {Johnson}, \citenamefont {Kleyheeg}, \citenamefont {Afach}, \citenamefont
  {Blanchard}, \citenamefont {Centers}, \citenamefont {Garcon}, \citenamefont
  {Engler}, \citenamefont {Figueroa}, \citenamefont {Sendra}, \citenamefont
  {Wickenbrock}, \citenamefont {Lawson}, \citenamefont {Wang}, \citenamefont
  {Wu}, \citenamefont {Luo}, \citenamefont {Mani}, \citenamefont {Mauskopf},
  \citenamefont {Graham}, \citenamefont {Rajendran}, \citenamefont {Kimball},
  \citenamefont {Budker},\ and\ \citenamefont {Sushkov}}]{Aybas2021a}%
  \BibitemOpen
  \bibfield  {author} {\bibinfo {author} {\bibfnamefont {D.}~\bibnamefont
  {Aybas}}, \bibinfo {author} {\bibfnamefont {J.}~\bibnamefont {Adam}},
  \bibinfo {author} {\bibfnamefont {E.}~\bibnamefont {Blumenthal}}, \bibinfo
  {author} {\bibfnamefont {A.~V.}\ \bibnamefont {Gramolin}}, \bibinfo {author}
  {\bibfnamefont {D.}~\bibnamefont {Johnson}}, \bibinfo {author} {\bibfnamefont
  {A.}~\bibnamefont {Kleyheeg}}, \bibinfo {author} {\bibfnamefont
  {S.}~\bibnamefont {Afach}}, \bibinfo {author} {\bibfnamefont {J.~W.}\
  \bibnamefont {Blanchard}}, \bibinfo {author} {\bibfnamefont {G.~P.}\
  \bibnamefont {Centers}}, \bibinfo {author} {\bibfnamefont {A.}~\bibnamefont
  {Garcon}}, \bibinfo {author} {\bibfnamefont {M.}~\bibnamefont {Engler}},
  \bibinfo {author} {\bibfnamefont {N.~L.}\ \bibnamefont {Figueroa}}, \bibinfo
  {author} {\bibfnamefont {M.~G.}\ \bibnamefont {Sendra}}, \bibinfo {author}
  {\bibfnamefont {A.}~\bibnamefont {Wickenbrock}}, \bibinfo {author}
  {\bibfnamefont {M.}~\bibnamefont {Lawson}}, \bibinfo {author} {\bibfnamefont
  {T.}~\bibnamefont {Wang}}, \bibinfo {author} {\bibfnamefont {T.}~\bibnamefont
  {Wu}}, \bibinfo {author} {\bibfnamefont {H.}~\bibnamefont {Luo}}, \bibinfo
  {author} {\bibfnamefont {H.}~\bibnamefont {Mani}}, \bibinfo {author}
  {\bibfnamefont {P.}~\bibnamefont {Mauskopf}}, \bibinfo {author}
  {\bibfnamefont {P.~W.}\ \bibnamefont {Graham}}, \bibinfo {author}
  {\bibfnamefont {S.}~\bibnamefont {Rajendran}}, \bibinfo {author}
  {\bibfnamefont {D.~F.}\ \bibnamefont {Kimball}}, \bibinfo {author}
  {\bibfnamefont {D.}~\bibnamefont {Budker}},\ and\ \bibinfo {author}
  {\bibfnamefont {A.~O.}\ \bibnamefont {Sushkov}},\ }\href
  {https://doi.org/10.1103/PhysRevLett.126.141802} {\bibfield  {journal}
  {\bibinfo  {journal} {Physical Review Letters}\ }\textbf {\bibinfo {volume}
  {126}},\ \bibinfo {pages} {141802} (\bibinfo {year} {2021}{\natexlab{a}})},\
  \Eprint {https://arxiv.org/abs/2101.01241} {arXiv:2101.01241} \BibitemShut
  {NoStop}%
\bibitem [{\citenamefont {Graham}\ and\ \citenamefont
  {Rajendran}(2013)}]{Graham2013}%
  \BibitemOpen
  \bibfield  {author} {\bibinfo {author} {\bibfnamefont {P.~W.}\ \bibnamefont
  {Graham}}\ and\ \bibinfo {author} {\bibfnamefont {S.}~\bibnamefont
  {Rajendran}},\ }\href {https://doi.org/10.1103/PhysRevD.88.035023} {\bibfield
   {journal} {\bibinfo  {journal} {Physical Review D - Particles, Fields,
  Gravitation and Cosmology}\ }\textbf {\bibinfo {volume} {88}},\ \bibinfo
  {pages} {035023} (\bibinfo {year} {2013})},\ \Eprint
  {https://arxiv.org/abs/1306.6088} {arXiv:1306.6088} \BibitemShut {NoStop}%
\bibitem [{\citenamefont {Laguta}\ \emph {et~al.}(2000)\citenamefont {Laguta},
  \citenamefont {Glinchuk}, \citenamefont {Slipenyuk},\ and\ \citenamefont
  {Bykov}}]{Laguta2000}%
  \BibitemOpen
  \bibfield  {author} {\bibinfo {author} {\bibfnamefont {V.~V.}\ \bibnamefont
  {Laguta}}, \bibinfo {author} {\bibfnamefont {M.~D.}\ \bibnamefont
  {Glinchuk}}, \bibinfo {author} {\bibfnamefont {A.~M.}\ \bibnamefont
  {Slipenyuk}},\ and\ \bibinfo {author} {\bibfnamefont {I.~P.}\ \bibnamefont
  {Bykov}},\ }\href {https://doi.org/10.1134/1.1332149} {\bibfield  {journal}
  {\bibinfo  {journal} {Physics of the Solid State}\ }\textbf {\bibinfo
  {volume} {42}},\ \bibinfo {pages} {2258} (\bibinfo {year}
  {2000})}\BibitemShut {NoStop}%
\bibitem [{\citenamefont {Warren}\ and\ \citenamefont
  {Robertson}(1996)}]{Warren1996}%
  \BibitemOpen
  \bibfield  {author} {\bibinfo {author} {\bibfnamefont {W.}~\bibnamefont
  {Warren}}\ and\ \bibinfo {author} {\bibfnamefont {J.}~\bibnamefont
  {Robertson}},\ }\href {https://doi.org/10.1103/PhysRevB.53.3080} {\bibfield
  {journal} {\bibinfo  {journal} {Physical Review B - Condensed Matter and
  Materials Physics}\ }\textbf {\bibinfo {volume} {53}},\ \bibinfo {pages}
  {3080} (\bibinfo {year} {1996})}\BibitemShut {NoStop}%
\bibitem [{\citenamefont {Warren}\ \emph {et~al.}(1993)\citenamefont {Warren},
  \citenamefont {Tuttle}, \citenamefont {McWhorter}, \citenamefont {Rong},\
  and\ \citenamefont {Poindexter}}]{Warren1993}%
  \BibitemOpen
  \bibfield  {author} {\bibinfo {author} {\bibfnamefont {W.~L.}\ \bibnamefont
  {Warren}}, \bibinfo {author} {\bibfnamefont {B.~A.}\ \bibnamefont {Tuttle}},
  \bibinfo {author} {\bibfnamefont {P.~J.}\ \bibnamefont {McWhorter}}, \bibinfo
  {author} {\bibfnamefont {F.~C.}\ \bibnamefont {Rong}},\ and\ \bibinfo
  {author} {\bibfnamefont {E.~H.}\ \bibnamefont {Poindexter}},\ }\href
  {https://doi.org/10.1063/1.108940} {\bibfield  {journal} {\bibinfo  {journal}
  {Applied Physics Letters}\ }\textbf {\bibinfo {volume} {62}},\ \bibinfo
  {pages} {482} (\bibinfo {year} {1993})}\BibitemShut {NoStop}%
\bibitem [{\citenamefont {Warren}\ \emph {et~al.}(1992)\citenamefont {Warren},
  \citenamefont {Seager}, \citenamefont {Dimos},\ and\ \citenamefont
  {Friebele}}]{Warren1992}%
  \BibitemOpen
  \bibfield  {author} {\bibinfo {author} {\bibfnamefont {W.~L.}\ \bibnamefont
  {Warren}}, \bibinfo {author} {\bibfnamefont {C.~H.}\ \bibnamefont {Seager}},
  \bibinfo {author} {\bibfnamefont {D.}~\bibnamefont {Dimos}},\ and\ \bibinfo
  {author} {\bibfnamefont {E.~J.}\ \bibnamefont {Friebele}},\ }\href
  {https://doi.org/10.1063/1.108171} {\bibfield  {journal} {\bibinfo  {journal}
  {Applied Physics Letters}\ }\textbf {\bibinfo {volume} {61}},\ \bibinfo
  {pages} {2530} (\bibinfo {year} {1992})}\BibitemShut {NoStop}%
\bibitem [{\citenamefont {{G{\'o}mez-Vidales}}\ \emph
  {et~al.}(2013)\citenamefont {{G{\'o}mez-Vidales}}, \citenamefont
  {{Granados-Oliveros}}, \citenamefont {{Nieto-Camacho}}, \citenamefont
  {{Reyes-Sol{\'i}s}},\ and\ \citenamefont
  {{Jim{\'e}nez-Estrada}}}]{Gomez-Vidales2013}%
  \BibitemOpen
  \bibfield  {author} {\bibinfo {author} {\bibfnamefont {V.}~\bibnamefont
  {{G{\'o}mez-Vidales}}}, \bibinfo {author} {\bibfnamefont {G.}~\bibnamefont
  {{Granados-Oliveros}}}, \bibinfo {author} {\bibfnamefont {A.}~\bibnamefont
  {{Nieto-Camacho}}}, \bibinfo {author} {\bibfnamefont {M.}~\bibnamefont
  {{Reyes-Sol{\'i}s}}},\ and\ \bibinfo {author} {\bibfnamefont
  {M.}~\bibnamefont {{Jim{\'e}nez-Estrada}}},\ }\href
  {https://doi.org/10.1039/C3RA42848F} {\bibfield  {journal} {\bibinfo
  {journal} {RSC Advances}\ }\textbf {\bibinfo {volume} {4}},\ \bibinfo {pages}
  {1371} (\bibinfo {year} {2013})}\BibitemShut {NoStop}%
\bibitem [{\citenamefont {Cooper}\ \emph {et~al.}(2010)\citenamefont {Cooper},
  \citenamefont {Dimitrijevic},\ and\ \citenamefont {Nadeau}}]{Cooper2010}%
  \BibitemOpen
  \bibfield  {author} {\bibinfo {author} {\bibfnamefont {D.~R.}\ \bibnamefont
  {Cooper}}, \bibinfo {author} {\bibfnamefont {N.~M.}\ \bibnamefont
  {Dimitrijevic}},\ and\ \bibinfo {author} {\bibfnamefont {J.~L.}\ \bibnamefont
  {Nadeau}},\ }\href {https://doi.org/10.1039/B9NR00130A} {\bibfield  {journal}
  {\bibinfo  {journal} {Nanoscale}\ }\textbf {\bibinfo {volume} {2}},\ \bibinfo
  {pages} {114} (\bibinfo {year} {2010})}\BibitemShut {NoStop}%
\bibitem [{\citenamefont {Sands}(1955)}]{Sands1955}%
  \BibitemOpen
  \bibfield  {author} {\bibinfo {author} {\bibfnamefont {R.~H.}\ \bibnamefont
  {Sands}},\ }\href {https://doi.org/10.1103/PhysRev.99.1222} {\bibfield
  {journal} {\bibinfo  {journal} {Physical Review}\ }\textbf {\bibinfo {volume}
  {99}},\ \bibinfo {pages} {1222} (\bibinfo {year} {1955})}\BibitemShut
  {NoStop}%
\bibitem [{\citenamefont {Abragam}\ and\ \citenamefont
  {Bleaney}(1970)}]{Abragam1970}%
  \BibitemOpen
  \bibfield  {author} {\bibinfo {author} {\bibfnamefont {A.}~\bibnamefont
  {Abragam}}\ and\ \bibinfo {author} {\bibfnamefont {B.}~\bibnamefont
  {Bleaney}},\ }\href@noop {} {\emph {\bibinfo {title} {Electron {{Paramagnetic
  Resonance}} of {{Transition Ions}}}}}\ (\bibinfo  {publisher} {Clarendon
  P.},\ \bibinfo {year} {1970})\ \Eprint {https://arxiv.org/abs/1011.1669v3}
  {arXiv:1011.1669v3} \BibitemShut {NoStop}%
\bibitem [{\citenamefont {Bales}\ \emph {et~al.}(1998)\citenamefont {Bales},
  \citenamefont {Peric},\ and\ \citenamefont {{Lamy-Freund}}}]{Bales1998}%
  \BibitemOpen
  \bibfield  {author} {\bibinfo {author} {\bibfnamefont {B.~L.}\ \bibnamefont
  {Bales}}, \bibinfo {author} {\bibfnamefont {M.}~\bibnamefont {Peric}},\ and\
  \bibinfo {author} {\bibfnamefont {M.~T.}\ \bibnamefont {{Lamy-Freund}}},\
  }\href {https://doi.org/10.1006/jmre.1998.1414} {\bibfield  {journal}
  {\bibinfo  {journal} {Journal of Magnetic Resonance}\ }\textbf {\bibinfo
  {volume} {132}},\ \bibinfo {pages} {279} (\bibinfo {year}
  {1998})}\BibitemShut {NoStop}%
\bibitem [{\citenamefont {Eaton}\ \emph {et~al.}(2010)\citenamefont {Eaton},
  \citenamefont {Eaton}, \citenamefont {Barr},\ and\ \citenamefont
  {Weber}}]{Eaton2010}%
  \BibitemOpen
  \bibfield  {author} {\bibinfo {author} {\bibfnamefont {G.~R.}\ \bibnamefont
  {Eaton}}, \bibinfo {author} {\bibfnamefont {S.~S.}\ \bibnamefont {Eaton}},
  \bibinfo {author} {\bibfnamefont {D.~P.}\ \bibnamefont {Barr}},\ and\
  \bibinfo {author} {\bibfnamefont {R.~T.}\ \bibnamefont {Weber}},\ }\href
  {https://doi.org/10.1007/978-3-211-92948-3} {\emph {\bibinfo {title}
  {Quantitative {{EPR}}}}}\ (\bibinfo  {publisher} {Springer},\ \bibinfo
  {address} {Vienna},\ \bibinfo {year} {2010})\BibitemShut {NoStop}%
\bibitem [{\citenamefont {Bairavarasu}\ \emph {et~al.}(2006)\citenamefont
  {Bairavarasu}, \citenamefont {Edwards}, \citenamefont {Sastry}, \citenamefont
  {Kochary}, \citenamefont {Lianos},\ and\ \citenamefont
  {Aggarwal}}]{Bairavarasu2006}%
  \BibitemOpen
  \bibfield  {author} {\bibinfo {author} {\bibfnamefont {S.}~\bibnamefont
  {Bairavarasu}}, \bibinfo {author} {\bibfnamefont {M.~E.}\ \bibnamefont
  {Edwards}}, \bibinfo {author} {\bibfnamefont {M.~D.}\ \bibnamefont {Sastry}},
  \bibinfo {author} {\bibfnamefont {F.}~\bibnamefont {Kochary}}, \bibinfo
  {author} {\bibfnamefont {D.}~\bibnamefont {Lianos}},\ and\ \bibinfo {author}
  {\bibfnamefont {M.~D.}\ \bibnamefont {Aggarwal}},\ }in\ \href
  {https://doi.org/10.1117/12.701456} {\emph {\bibinfo {booktitle}
  {Photorefractive {{Fiber}} and {{Crystal Devices}}: {{Materials}}, {{Optical
  Properties}}, and {{Applications XII}}}}},\ Vol.\ \bibinfo {volume} {6314}\
  (\bibinfo  {publisher} {SPIE},\ \bibinfo {year} {2006})\ pp.\ \bibinfo
  {pages} {42--48}\BibitemShut {NoStop}%
\bibitem [{\citenamefont {Waugh}\ and\ \citenamefont
  {Slichter}(1988)}]{Waugh1988}%
  \BibitemOpen
  \bibfield  {author} {\bibinfo {author} {\bibfnamefont {J.~S.}\ \bibnamefont
  {Waugh}}\ and\ \bibinfo {author} {\bibfnamefont {C.~P.}\ \bibnamefont
  {Slichter}},\ }\href {https://doi.org/10.1103/PhysRevB.37.4337} {\bibfield
  {journal} {\bibinfo  {journal} {Physical Review B}\ }\textbf {\bibinfo
  {volume} {37}},\ \bibinfo {pages} {4337} (\bibinfo {year}
  {1988})}\BibitemShut {NoStop}%
\bibitem [{\citenamefont {Khutsishvili}(1969)}]{Khutsishvili1969}%
  \BibitemOpen
  \bibfield  {author} {\bibinfo {author} {\bibfnamefont {G.~R.}\ \bibnamefont
  {Khutsishvili}},\ }\href@noop {} {\bibfield  {journal} {\bibinfo  {journal}
  {Physics-Uspekhi}\ }\textbf {\bibinfo {volume} {11}},\ \bibinfo {pages} {802}
  (\bibinfo {year} {1969})}\BibitemShut {NoStop}%
\bibitem [{\citenamefont {Henrichs}\ \emph {et~al.}(1984)\citenamefont
  {Henrichs}, \citenamefont {Cofield}, \citenamefont {Young},\ and\
  \citenamefont {Michael~Hewitt}}]{Henrichs1984}%
  \BibitemOpen
  \bibfield  {author} {\bibinfo {author} {\bibfnamefont {P.~M.}\ \bibnamefont
  {Henrichs}}, \bibinfo {author} {\bibfnamefont {M.~L.}\ \bibnamefont
  {Cofield}}, \bibinfo {author} {\bibfnamefont {R.~H.}\ \bibnamefont {Young}},\
  and\ \bibinfo {author} {\bibfnamefont {J.}~\bibnamefont {Michael~Hewitt}},\
  }\href {https://doi.org/10.1016/0022-2364(84)90009-X} {\bibfield  {journal}
  {\bibinfo  {journal} {Journal of Magnetic Resonance (1969)}\ }\textbf
  {\bibinfo {volume} {58}},\ \bibinfo {pages} {85} (\bibinfo {year}
  {1984})}\BibitemShut {NoStop}%
\bibitem [{\citenamefont {Adam}(2023)}]{Adam2023}%
  \BibitemOpen
  \bibfield  {author} {\bibinfo {author} {\bibfnamefont {J.}~\bibnamefont
  {Adam}},\ }\emph {\bibinfo {title} {Search for {{Axion Dark Matter Using
  Solid State Nuclear Magnetic Resonance}} and {{Superconducting
  Magnetometers}}}},\ \href@noop {} {Ph.D. thesis},\ \bibinfo  {school}
  {ProQuest Dissertations \& Theses} (\bibinfo {year} {2023})\BibitemShut
  {NoStop}%
\bibitem [{\citenamefont {Mukhamedjanov}\ and\ \citenamefont
  {Sushkov}(2005)}]{Mukhamedjanov2005a}%
  \BibitemOpen
  \bibfield  {author} {\bibinfo {author} {\bibfnamefont {T.~N.}\ \bibnamefont
  {Mukhamedjanov}}\ and\ \bibinfo {author} {\bibfnamefont {O.~P.}\ \bibnamefont
  {Sushkov}},\ }\href {https://doi.org/10.1103/PhysRevA.72.034501} {\bibfield
  {journal} {\bibinfo  {journal} {Physical Review A}\ }\textbf {\bibinfo
  {volume} {72}},\ \bibinfo {pages} {34501} (\bibinfo {year} {2005})},\ \Eprint
  {https://arxiv.org/abs/physics/0411226} {arXiv:physics/0411226} \BibitemShut
  {NoStop}%
\bibitem [{\citenamefont {Budker}\ \emph {et~al.}(2006)\citenamefont {Budker},
  \citenamefont {Lamoreaux}, \citenamefont {Sushkov},\ and\ \citenamefont
  {Sushkov}}]{Budker2006}%
  \BibitemOpen
  \bibfield  {author} {\bibinfo {author} {\bibfnamefont {D.}~\bibnamefont
  {Budker}}, \bibinfo {author} {\bibfnamefont {S.~K.}\ \bibnamefont
  {Lamoreaux}}, \bibinfo {author} {\bibfnamefont {A.~O.}\ \bibnamefont
  {Sushkov}},\ and\ \bibinfo {author} {\bibfnamefont {O.~P.}\ \bibnamefont
  {Sushkov}},\ }\href {https://doi.org/10.1103/PhysRevA.73.022107} {\bibfield
  {journal} {\bibinfo  {journal} {Physical Review A}\ }\textbf {\bibinfo
  {volume} {73}},\ \bibinfo {pages} {022107} (\bibinfo {year}
  {2006})}\BibitemShut {NoStop}%
\bibitem [{\citenamefont {DeMille}\ \emph {et~al.}(2017)\citenamefont
  {DeMille}, \citenamefont {Doyle},\ and\ \citenamefont
  {Sushkov}}]{DeMille2017}%
  \BibitemOpen
  \bibfield  {author} {\bibinfo {author} {\bibfnamefont {D.}~\bibnamefont
  {DeMille}}, \bibinfo {author} {\bibfnamefont {J.~M.}\ \bibnamefont {Doyle}},\
  and\ \bibinfo {author} {\bibfnamefont {A.~O.}\ \bibnamefont {Sushkov}},\
  }\href {https://doi.org/10.1126/science.aal3003} {\bibfield  {journal}
  {\bibinfo  {journal} {Science}\ }\textbf {\bibinfo {volume} {357}},\ \bibinfo
  {pages} {990} (\bibinfo {year} {2017})}\BibitemShut {NoStop}%
\bibitem [{\citenamefont {Wu}\ \emph {et~al.}(2019)\citenamefont {Wu},
  \citenamefont {Blanchard}, \citenamefont {Centers}, \citenamefont {Figueroa},
  \citenamefont {Garcon}, \citenamefont {Graham}, \citenamefont {Kimball},
  \citenamefont {Rajendran}, \citenamefont {Stadnik}, \citenamefont {Sushkov},
  \citenamefont {Wickenbrock},\ and\ \citenamefont {Budker}}]{Wu2019a}%
  \BibitemOpen
  \bibfield  {author} {\bibinfo {author} {\bibfnamefont {T.}~\bibnamefont
  {Wu}}, \bibinfo {author} {\bibfnamefont {J.~W.}\ \bibnamefont {Blanchard}},
  \bibinfo {author} {\bibfnamefont {G.~P.}\ \bibnamefont {Centers}}, \bibinfo
  {author} {\bibfnamefont {N.~L.}\ \bibnamefont {Figueroa}}, \bibinfo {author}
  {\bibfnamefont {A.}~\bibnamefont {Garcon}}, \bibinfo {author} {\bibfnamefont
  {P.~W.}\ \bibnamefont {Graham}}, \bibinfo {author} {\bibfnamefont {D.~F.}\
  \bibnamefont {Kimball}}, \bibinfo {author} {\bibfnamefont {S.}~\bibnamefont
  {Rajendran}}, \bibinfo {author} {\bibfnamefont {Y.~V.}\ \bibnamefont
  {Stadnik}}, \bibinfo {author} {\bibfnamefont {A.~O.}\ \bibnamefont
  {Sushkov}}, \bibinfo {author} {\bibfnamefont {A.}~\bibnamefont
  {Wickenbrock}},\ and\ \bibinfo {author} {\bibfnamefont {D.}~\bibnamefont
  {Budker}},\ }\href {https://doi.org/10.1103/PhysRevLett.122.191302}
  {\bibfield  {journal} {\bibinfo  {journal} {Physical Review Letters}\
  }\textbf {\bibinfo {volume} {122}},\ \bibinfo {pages} {191302} (\bibinfo
  {year} {2019})},\ \Eprint {https://arxiv.org/abs/1901.10843}
  {arXiv:1901.10843} \BibitemShut {NoStop}%
\bibitem [{\citenamefont {Garcon}\ \emph {et~al.}(2019)\citenamefont {Garcon},
  \citenamefont {Blanchard}, \citenamefont {Centers}, \citenamefont {Figueroa},
  \citenamefont {Graham}, \citenamefont {Jackson~Kimball}, \citenamefont
  {Rajendran}, \citenamefont {Sushkov}, \citenamefont {Stadnik}, \citenamefont
  {Wickenbrock}, \citenamefont {Wu},\ and\ \citenamefont
  {Budker}}]{Garcon2019}%
  \BibitemOpen
  \bibfield  {author} {\bibinfo {author} {\bibfnamefont {A.}~\bibnamefont
  {Garcon}}, \bibinfo {author} {\bibfnamefont {J.~W.}\ \bibnamefont
  {Blanchard}}, \bibinfo {author} {\bibfnamefont {G.~P.}\ \bibnamefont
  {Centers}}, \bibinfo {author} {\bibfnamefont {N.~L.}\ \bibnamefont
  {Figueroa}}, \bibinfo {author} {\bibfnamefont {P.~W.}\ \bibnamefont
  {Graham}}, \bibinfo {author} {\bibfnamefont {D.~F.}\ \bibnamefont
  {Jackson~Kimball}}, \bibinfo {author} {\bibfnamefont {S.}~\bibnamefont
  {Rajendran}}, \bibinfo {author} {\bibfnamefont {A.~O.}\ \bibnamefont
  {Sushkov}}, \bibinfo {author} {\bibfnamefont {Y.~V.}\ \bibnamefont
  {Stadnik}}, \bibinfo {author} {\bibfnamefont {A.}~\bibnamefont
  {Wickenbrock}}, \bibinfo {author} {\bibfnamefont {T.}~\bibnamefont {Wu}},\
  and\ \bibinfo {author} {\bibfnamefont {D.}~\bibnamefont {Budker}},\ }\href
  {https://doi.org/10.1126/sciadv.aax4539} {\bibfield  {journal} {\bibinfo
  {journal} {Science Advances}\ }\textbf {\bibinfo {volume} {5}},\ \bibinfo
  {pages} {eaax4539} (\bibinfo {year} {2019})},\ \Eprint
  {https://arxiv.org/abs/1902.04644} {arXiv:1902.04644} \BibitemShut {NoStop}%
\bibitem [{\citenamefont {Aybas}\ \emph
  {et~al.}(2021{\natexlab{b}})\citenamefont {Aybas}, \citenamefont {Bekker},
  \citenamefont {Blanchard}, \citenamefont {Budker}, \citenamefont {Centers},
  \citenamefont {Figueroa}, \citenamefont {Gramolin}, \citenamefont
  {Jackson~Kimball}, \citenamefont {Wickenbrock},\ and\ \citenamefont
  {Sushkov}}]{Aybas2021b}%
  \BibitemOpen
  \bibfield  {author} {\bibinfo {author} {\bibfnamefont {D.}~\bibnamefont
  {Aybas}}, \bibinfo {author} {\bibfnamefont {H.}~\bibnamefont {Bekker}},
  \bibinfo {author} {\bibfnamefont {J.~W.}\ \bibnamefont {Blanchard}}, \bibinfo
  {author} {\bibfnamefont {D.}~\bibnamefont {Budker}}, \bibinfo {author}
  {\bibfnamefont {G.~P.}\ \bibnamefont {Centers}}, \bibinfo {author}
  {\bibfnamefont {N.~L.}\ \bibnamefont {Figueroa}}, \bibinfo {author}
  {\bibfnamefont {A.~V.}\ \bibnamefont {Gramolin}}, \bibinfo {author}
  {\bibfnamefont {D.~F.}\ \bibnamefont {Jackson~Kimball}}, \bibinfo {author}
  {\bibfnamefont {A.}~\bibnamefont {Wickenbrock}},\ and\ \bibinfo {author}
  {\bibfnamefont {A.~O.}\ \bibnamefont {Sushkov}},\ }\href
  {https://doi.org/10.1088/2058-9565/abfbbc} {\bibfield  {journal} {\bibinfo
  {journal} {Quantum Science and Technology}\ }\textbf {\bibinfo {volume}
  {6}},\ \bibinfo {pages} {034007} (\bibinfo {year} {2021}{\natexlab{b}})},\
  \Eprint {https://arxiv.org/abs/2103.06284} {arXiv:2103.06284} \BibitemShut
  {NoStop}%
\bibitem [{\citenamefont {Sushkov}\ \emph {et~al.}(2023)\citenamefont
  {Sushkov}, \citenamefont {Sushkov},\ and\ \citenamefont
  {Yaresko}}]{Sushkov2023}%
  \BibitemOpen
  \bibfield  {author} {\bibinfo {author} {\bibfnamefont {A.~O.}\ \bibnamefont
  {Sushkov}}, \bibinfo {author} {\bibfnamefont {O.~P.}\ \bibnamefont
  {Sushkov}},\ and\ \bibinfo {author} {\bibfnamefont {A.}~\bibnamefont
  {Yaresko}},\ }\href {https://doi.org/10.1103/PhysRevA.107.062823} {\bibfield
  {journal} {\bibinfo  {journal} {Physical Review A}\ }\textbf {\bibinfo
  {volume} {107}},\ \bibinfo {pages} {062823} (\bibinfo {year}
  {2023})}\BibitemShut {NoStop}%
\bibitem [{\citenamefont {Portis}(1953)}]{Portis1953}%
  \BibitemOpen
  \bibfield  {author} {\bibinfo {author} {\bibfnamefont {A.~M.}\ \bibnamefont
  {Portis}},\ }\href {https://doi.org/10.1103/PhysRev.91.1071} {\bibfield
  {journal} {\bibinfo  {journal} {Physical Review}\ }\textbf {\bibinfo {volume}
  {91}},\ \bibinfo {pages} {1071} (\bibinfo {year} {1953})}\BibitemShut
  {NoStop}%
\bibitem [{\citenamefont {Wang}\ \emph {et~al.}(2014)\citenamefont {Wang},
  \citenamefont {Wang}, \citenamefont {Ge}, \citenamefont {Luo}, \citenamefont
  {Li}, \citenamefont {Viehland}, \citenamefont {Chen},\ and\ \citenamefont
  {Luo}}]{Wang2014a}%
  \BibitemOpen
  \bibfield  {author} {\bibinfo {author} {\bibfnamefont {Y.}~\bibnamefont
  {Wang}}, \bibinfo {author} {\bibfnamefont {Z.}~\bibnamefont {Wang}}, \bibinfo
  {author} {\bibfnamefont {W.}~\bibnamefont {Ge}}, \bibinfo {author}
  {\bibfnamefont {C.}~\bibnamefont {Luo}}, \bibinfo {author} {\bibfnamefont
  {J.}~\bibnamefont {Li}}, \bibinfo {author} {\bibfnamefont {D.}~\bibnamefont
  {Viehland}}, \bibinfo {author} {\bibfnamefont {J.}~\bibnamefont {Chen}},\
  and\ \bibinfo {author} {\bibfnamefont {H.}~\bibnamefont {Luo}},\ }\href
  {https://doi.org/10.1103/PhysRevB.90.134107} {\bibfield  {journal} {\bibinfo
  {journal} {Physical Review B}\ }\textbf {\bibinfo {volume} {90}},\ \bibinfo
  {pages} {134107} (\bibinfo {year} {2014})}\BibitemShut {NoStop}%
\bibitem [{\citenamefont {More}\ \emph {et~al.}(1984)\citenamefont {More},
  \citenamefont {Eaton},\ and\ \citenamefont {Eaton}}]{More1984}%
  \BibitemOpen
  \bibfield  {author} {\bibinfo {author} {\bibfnamefont {K.~M.}\ \bibnamefont
  {More}}, \bibinfo {author} {\bibfnamefont {G.~R.}\ \bibnamefont {Eaton}},\
  and\ \bibinfo {author} {\bibfnamefont {S.~S.}\ \bibnamefont {Eaton}},\ }\href
  {https://doi.org/10.1016/0022-2364(84)90025-8} {\bibfield  {journal}
  {\bibinfo  {journal} {Journal of Magnetic Resonance (1969)}\ }\textbf
  {\bibinfo {volume} {60}},\ \bibinfo {pages} {54} (\bibinfo {year}
  {1984})}\BibitemShut {NoStop}%
\bibitem [{\citenamefont {Dalal}\ \emph {et~al.}(1981)\citenamefont {Dalal},
  \citenamefont {Eaton},\ and\ \citenamefont {Eaton}}]{Dalal1981}%
  \BibitemOpen
  \bibfield  {author} {\bibinfo {author} {\bibfnamefont {D.~P.}\ \bibnamefont
  {Dalal}}, \bibinfo {author} {\bibfnamefont {S.~S.}\ \bibnamefont {Eaton}},\
  and\ \bibinfo {author} {\bibfnamefont {G.~R.}\ \bibnamefont {Eaton}},\ }\href
  {https://doi.org/10.1016/0022-2364(81)90276-6} {\bibfield  {journal}
  {\bibinfo  {journal} {Journal of Magnetic Resonance (1969)}\ }\textbf
  {\bibinfo {volume} {44}},\ \bibinfo {pages} {415} (\bibinfo {year}
  {1981})}\BibitemShut {NoStop}%
\bibitem [{\citenamefont {Schreurs}\ and\ \citenamefont
  {Fraenkel}(1961)}]{Schreurs1961}%
  \BibitemOpen
  \bibfield  {author} {\bibinfo {author} {\bibfnamefont {J.~W.~H.}\
  \bibnamefont {Schreurs}}\ and\ \bibinfo {author} {\bibfnamefont {G.~K.}\
  \bibnamefont {Fraenkel}},\ }\href {https://doi.org/10.1063/1.1731672}
  {\bibfield  {journal} {\bibinfo  {journal} {The Journal of Chemical Physics}\
  }\textbf {\bibinfo {volume} {34}},\ \bibinfo {pages} {756} (\bibinfo {year}
  {1961})}\BibitemShut {NoStop}%
\end{thebibliography}%

\end{document}